\documentclass[emulateapj,twocolumn,tighten,times]{aastex61}

\newcommand\icgc{ICGC}
\newcommand\icgcs{ICGCs}
\newcommand\sx{{\sc SExtractor}~}

\newcommand\galf{{\sc GALFIT}~}
\newcommand\galfit{{\sc GALFIT}}
\newcommand\ser{S\'ersic~}
\newcommand\sersic{S\'ersic}
\newcommand\cser{core-S\'ersic~}
\newcommand\csersic{core-S\'ersic}
\newcommand\cote{C{\^ o}t{\' e}}

\newcommand\hst{{\it HST}}
\newcommand\jwst{{\it JWST}}
\newcommand\etal{{~et\,al.~}}

\newcommand\ngc{\ensuremath{N_{\rm GC}}}
\newcommand\Mbh{\ensuremath{M_{\rm BH}}}

\newcommand\Iacs{\ensuremath{I_{814}}}
\newcommand\iacs{\ensuremath{i_{775}}}

\newcommand\mM{\ensuremath{(m{-}M)}}
\newcommand\mo{$M_{\odot}$}

\shorttitle{Specific frequencies of galaxies and ICL in A1689}
\shortauthors{Alamo-Mart\'inez \& Blakeslee}
\submitjournal{ApJ}
\accepted{September 24, 2017}
\begin{document}

\title{Specific Frequencies and Luminosity Profiles of Cluster Galaxies \\
 and Intracluster Light in Abell 1689}

\correspondingauthor{K. A. Alamo-Mart\'{i}nez}
\email{kalamo@astro.puc.cl}

\author[0000-0002-5897-7813]{K. A. Alamo-Mart\'{i}nez}
\affil{Instituto de Astrof\'isica, Pontificia Universidad Cat\'olica de Chile, 7820436
  Macul, Santiago, Chile}

\author[0000-0002-5213-3548]{J. P. Blakeslee}
\affil{Herzberg Astronomy \& Astrophysics, National Research Council of Canada, Victoria, BC V9E 2E7, Canada}

\begin{abstract}\noindent
We present magnitudes and profile fits for 180 galaxies in the central
field of the massive lensing cluster Abell~1689 using very deep imaging 
with the \textit{Hubble Space Telescope} Advanced Camera for Surveys in the F814W bandpass.
Previous work revealed an exceptionally large number of globular clusters (GCs) 
in A1689 and mapped their number density distribution. We decompose this
number density map into GCs associated with individual cluster galaxies and
ICGCs (intracluster globular clusters) associated with the intracluster light (ICL).
In all, we measure GC specific frequencies $S_N$ for 33 cluster members and the ICL.
The relation between $S_N$ and galaxy magnitude is
consistent with the trend observed in Virgo, although some intermediate
luminosity galaxies scatter to $S_N>10$.  We estimate the ICL makes up 11\% of the
starlight in this field, whereas the ICGCs account for $\sim\,$35\% of the GCs, both
consistent with predictions from simulations.  Galaxies with higher $S_N$ values tend
to be rounder, and there is a marginally significant trend of decreasing $S_N$ with
increasing specific angular momenta~$\lambda_R$. 
We also reevaluate the GC population in the A2744 Frontier Field, for which fewer than one-tenth as many GCs have been detected because of its larger distance. 
Finally, our \csersic\ fit to the light profile of the A1689 BCG implies a break radius of 3.8~kpc, among the largest
known; we discuss implications of the sizable core and extensive GC
population for the supermassive black hole in light of scaling relations.
\end{abstract}

%===========================================================
%=====================   INTRODUCTION   ======================
%===========================================================

\section{Introduction}

%It has been known for more than half a century that the morphological and
It has been known for the better part of a century that the morphological and
spectral properties of galaxies correlate with the
surrounding environment (e.g., Spitzer \& Baade 1951;
Zwicky \& Humason 1964; Abell 1965; Einasto et al.\ 1974; Oemler 1974).
In the intervening years, numerous studies have used large observational data sets
or numerical simulations to explore and quantify how the baryonic
properties of galaxies are molded by their surroundings (e.g., Dressler 1980;
Kauffmann\etal2004; Postman et al.\ 2005; Tanaka\etal2005; Baldry et al.\ 2006;
Chung\etal2009; Vulcani\etal2012; Jaffee\etal2016).
Interactions among galaxies and with the prevailing group or cluster medium, as well as
the cessation of cold  gas accretion, play important roles in  the evolution of galaxies
within dense regions (e.g., van den  Bosch et al.\ 2008;  van der Wel\etal2010;
Peng\etal2010; Muzzin\etal2012; Mok\etal2013; Smethurst\etal2017).
However, many studies suggest such ``environmental quenching'' mechanisms play a
secondary role and are mainly applicable to satellite galaxies, while ``self quenching''
(e.g., via AGN feedback or other internal processes) of individual galaxies, regulated
by halo mass, is the main driver in the evolution for central galaxies (e.g., Mandelbaum
et al.\ 2006; Peng et al.\ 2012; Wetzel et al.\ 2013; Tal et al.\ 2014; Bluck et
al.\ 2016).  Still others have argued that these two broad varieties of quenching are no
longer distinct for massive galaxies within cluster-sized halos (Knobel et al.\ 2015);
moreover, the diverse underlying physical processes vary with redshift (Balogh et
al.\ 2016; Fossati et al.\ 2017)

Regardless of the precise balance between environmental and mass-regulated quenching,
it is clear that large galaxies today are complex systems, assembled over
cosmic time through the merging of multiple stellar systems, gas accretion, and in situ star
formation, all subjected to a variety of external and internal transformative processes.
These processes are accentuated among the most massive galaxies and within the densest
environments; therefore, an arguably good place to study them,
or at least their end products, are the massive early-type galaxies
that populate the centers of rich clusters.
Most of these objects have long ago consumed or lost most of their gas
and have settled down to passive evolution along the red sequence (see references
above), concealing most of the complex processes that shaped their formation.
However, a window into the ancient histories of cluster ellipticals is provided by 
the abundant populations of globular clusters (GCs) that they retain.
Compared to the stellar halos of the galaxies themselves, GCs are relatively
simple, old, mainly metal-poor systems that occur in
large numbers around early-type cluster galaxies (e.g., Harris 1991; Peng et al.\ 2008). 
The surviving GCs are very dense stellar systems that were formed at early epochs and
have been widely used as tracers of galactic structure
(see reviews by West\etal2004; Brodie \& Strader 2006).

The richness of a GC population is parameterized by its the specific frequency $S_N$
(Harris \& van den Bergh 1981), the number of GCs per unit $V$-band galaxy
luminosity. Among early-type galaxies, $S_N$ varies with the absolute magnitude of
the galaxy $M_V$ following a \textsf{U}-shape trend: it
increases with luminosity on the bright side, reaches a minimum at
luminosities $\sim L^*_V$, and then increases as luminosity decreases on
the faint side, though with a large dispersion (Harris 2001; Peng\etal2008).
The nonlinear behavior of $S_N$ may be understood if the number of GCs
(or the mass contained in them) scales with the total halo mass of the host galaxy
(Blakeslee et al.\ 1997; Blakeslee 1999; Kravtsov \& Gnedin 2005;
Spitler \& Forbes 2009; Hudson\etal2014; Harris\etal2017),
suggesting that $S_N$ scales inversely with the stellar-to-halo mass relation
(Moster et al.\ 2010; Hudson\etal2015), or the total star formation efficiency.
Peng\etal(2008) also found evidence for an environmental effect within Virgo,
whereby the majority of dwarf ellipticals in Virgo with high $S_N$ are located within
1\,Mpc of the cD galaxy M87. Moreover, Liu\etal(2016) 
measured [$\alpha$/Fe] for 11 of the faintest galaxies used in Peng\etal(2008), finding a tendency for $S_N$ to increase with [$\alpha$/Fe]. 
On the other hand, Georgiev\etal(2010) compiled a large sample of $S_N$ for
galaxies of varying morphologies in diverse environments and found no significant
environmental trends overall.

Another important component of galaxy clusters is the intracluster light
(ICL), diffuse stellar material gravitationally bound to the cluster potential
rather than individual galaxies, resulting from tidal stripping and relaxation
processes (Gallagher \& Ostriker 1972).
Numerical simulations and observational evidence suggest that 
the ICL comprises 10\%-40\% of the optical light in the cluster (Purcell\etal2007;
Rudick\etal2011; Contini\etal2014), although it is often challenging to distinguish this
material from extended structures of cluster members.
In addition to ICL, tidal interactions must also strip GCs from their parent galaxies,
producing a population of 
intracluster GCs (ICGCs) that are bound to the cluster as a whole (White 1987).
Based on estimated distances to several GCs found in the Palomar Observatory Sky
Survey, van den Bergh (1958) reported the discovery of  
several intergalactic GCs (the equivalent of ICGCs outside of clusters) in the
Local Group.  However, these objects are now generally considered distant members of
the Galactic GC system, and the existence of truly intergalactic GCs in the Local Group
remains in dispute (Mackey et al.\ 2016).

West\etal(1995) proposed large numbers of ICGCs could be responsible for the high values
of $S_N$ found for many brightest cluster galaxies (BCGs); these high-$S_N$ BCGs
would have additional GCs associated to them because of their privileged positions
within cluster centers.  Interestingly, Abell~1185, a cluster in which the central
galaxy is offset from the centroid of the X-ray emission, appears to have a significant
population of ICGCs located at the X-ray center of the cluster (Jord\'an\etal2003;
West\etal2011).  Peng\etal(2011) reported a large number of ICGCs in the Coma cluster,
comprising $\sim$30\%-45\% of the total population in the core, although with high
uncertainties due to the low fraction of the covered area. Evidence also exists for a
more modest number of ICGCs in the nearby Virgo cluster (Durrell et al.\ 2014; Ko et
al.\ 2017).  As is the case for the ICL, it is often difficult to distinguish ICGCs from
the outer members of the GC systems of individual cluster galaxies.  
However, sufficient observational evidence now exists to make
ICGCs an established feature of galaxy clusters.

%Also, they were not able to measure the ICL, but claimed that the ICL might have high
%$S_N$ values, similar to the high-$S_N$ BGC galaxies. A significant population of ICGCs
%resides in Abell~1185 (Jord\'an et al.\ 2003; West et~al.\ 2011), a cluster in which
%the cD galaxy is offset from the centroid of X-ray emission.

In the present work, we explore the luminosity profiles and specific frequencies of
early-type galaxies near the center of Abell\,1689 (A1689), an extremely massive galaxy
cluster with a very high central density of galaxies.  This is a follow-up of our
previous study of the GC population in A1689 (Alamo-Mart\'{i}nez\etal2013, hereafter AM13) where the
analysis was done for the cluster as whole, not for individual galaxies.  In that work
we constrained the total population of GCs within a projected radius of 400~kpc of the
center of A1689, and found it to be the most populous GC system known.
In this follow-up study, we decompose the stellar light and GC populations into
components representing individual galaxies, derive $S_N$ values for $\sim30$ members,
and constrain the ICL and number of ICGCs.
The following section summarizes the observational data and reductions.  Section~3
describes the modeling of the galaxy profiles, GC populations, and ICL.  In Section 4,
we present the resulting specific frequencies, explore a possible correlation with
kinematics from the literature, examine the possibility of offsets between the galaxy
light and GC systems, and discuss a lower mass estimate for the black hole
in the A1689 cD galaxy.  Section~5 presents our conclusions.
For consistency with our AM13 study, we adopt a cosmology
with $(h,\Omega_m,\Omega_{\lambda}) = (0.70,0.27,0.73)$, giving
a distance modulus for A1689 ($z{\,=\,}0.183$)
of $\mM = 39.738$ mag and a physical scale of 3.07~kpc~arcsec$^{-1}$.
All magnitudes are on the AB system except where noted otherwise.

%===========================================================
%================   DATA / OBSERVATIONS  ================
%===========================================================
\section{Data}
\label{data.sect}

\subsection{\textit{Hubble} Observation Summary}
\label{observations.sect}

The current work uses the same deep imaging data, obtained in Program GO-11710
with the \textit{Hubble Space Telescope} (\hst) Advanced Camera for Surveys
Wide Field Channel (ACS/WFC), as was analyzed by AM13.
To summarize, we observed the central region of A1689
for 28 orbits in the F814W bandpass, the ACS/WFC bandpass with the highest total throughput.
The individual exposures were
calibrated with the standard \hst\ pipeline, corrected for charge transfer inefficiency
(Anderson \& Bedin 2010),
and then geometrically corrected and combined with the Apsis pipeline
(Blakeslee\etal2003) into a single stacked image of total exposure time 75,172\,s and a
resampled pixel scale of 0\farcs033~pix$^{-1}$.
The point spread function (PSF) in the final image has a full width at half maximum
(FWHM) of 0\farcs086, or 2.6~pix, and thus is adequately sampled.
% (left panel of Figure~\ref{galfitModels}).
As in AM13, we use the AB photometric system, on which the F814W zero point is
25.947~mag; the Galactic extinction towards A1689 in this bandpass is 0.04\,mag
(Schlafly \& Finkbeiner 2011).  
We refer the reader to AM13 for a more detailed description of the observations 
and image reductions. 

\newpage
\subsection{Globular cluster density map}
\label{data_gc_map.sect}

At the distance of A1689, the GCs appear as point sources. Thus, in AM13 we optimized the source
detection and magnitude measurements for point sources.  The source detection was done
with \sx (Bertin \& Arnouts 1996) on a residual image with all large galaxies subtracted, 
bright stars and areas of poor subtraction masked, and a ring median filter applied;
magnitudes were derived from PSF photometry using DAOPhot (Stetson
1987) on an unfiltered version of the residual image at the coordinates of the
SExtractor detections.  We performed detailed, realistic completeness
tests, where the input magnitudes of the artificial stars followed a Gaussian
distribution modeled on the expected globular cluster luminosity function (GCLF).
The fraction of recovered stars as function of magnitude was described by
a four-parameter modified-Fermi function that included the effect of Eddington bias 
on the magnitudes near the detection limit (see AM13 for details).  Furthermore,
we corrected for contamination from background galaxies and foreground
stars and included the effect of gravitational lensing on the spatial and magnitude
distributions of the background galaxies.

In AM13, we detected a total of 8283 unresolved sources (classified as point sources by
DAOPhot sharpness and goodness-of-fit criteria) within the ACS/WFC field of view (FOV)
over the magnitude range $27.0<\Iacs<29.32$ mag, where the faint limit corresponds to the
magnitude where we recovered 50\% of the sources in the completeness tests, and the
bright limit corresponds to $\sim3\,\sigma$ brighter than the turnover of the GCLF.
(Note that the value of 8212 point source detections quoted by AM13 over this magnitude
range referred only to objects within the FOV and inside
a projected radius of 400~kpc of the cD galaxy;
see Fig.~3 of that work).
Using the coordinates of the objects in this sample, we constructed a surface number
density map of GC candidates, corrected for incompleteness and background contamination
as detailed in AM13, by smoothing with a Gaussian filter of FWHM = 10\arcsec\
($\sigma=4\farcs25$).  Masked regions were interpolated; following these corrections,
the integral of the GC density map over the FOV gives 7940 objects to the detection
limit. This represents just over 5\% of the GCLF; the correction for
integrating over the GCLF is a factor of $\sim\,$19.3.
AM13 mainly used the GC map for illustrative purposes in comparing to the galaxy light,
lensing mass, and X-ray surface brightness distributions.  In the present work, we use
this number density map to model the GC populations of the individual galaxies and ICL
component in A1689.

%===========================================================
%===========================================================
%========================   ANALYSIS   ========================
%===========================================================
%===========================================================

\begin{figure}
\centering
\includegraphics[angle=0,width=0.5\textwidth]{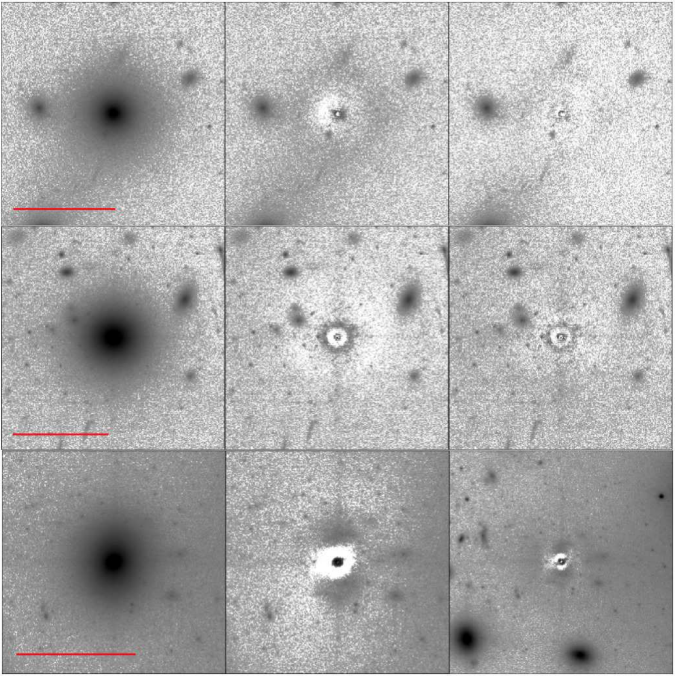}
\caption{Examples of single versus double \sersic\ model fit residuals.
  {Left panels:} examples of three typical A1689 galaxies in the ACS/WFC FOV. The red bar indicates the scale corresponding to 5~arcseconds. 
  {Middle panels:} residuals after subtraction of the best-fit single \sersic\ models.
  {Right panels:} residuals after subtraction of the best-fit double \sersic\ models.
\label{sersicFits_1c_vs_2c}}
\end{figure}

\section{Analysis}
\label{analysis.sect}

%========================   Modeling of individual galaxies   ========================
\subsection{Modeling of individual galaxies} 
\label{galaxymodel.sect}

A simple way to characterize quantitatively the structure of any galaxy is through
parametric fitting of its luminosity profile, 
and a \ser function (\ser 1968) is one of the most widely used models for
describing diverse types of galaxies. It has the form:
\begin{equation}
I(r)=I_{e}\exp\left\{-b_{n} \left[ {\left( \frac{r}{R_{e}}\right) }^{1/n}- 1\right] \right\}\ ,
\label{eq_intSersic}
\end{equation} 
\noindent
where $R_{e}$ is the effective radius that encloses half of the light; $I_{e}$
is the intensity at $R_{e}$; $n$ is the \ser index, which governs the shape of
the radial profile; and $b_{n}\approx1.9992n - 0.3271$ (Graham \& Driver 2005).  
To decompose the contributions of individual galaxies to the total luminosity profile in
A1689, we use \galf (Peng et al. 2010) to fit 2-D \ser functions, convolved with the PSF,
to 180 of the brightest galaxies in the ACS/WFC FOV.
%% give this info in next paragraph....
%  from which 94 are cluster members confirmed with spectroscopic or photometric
%  redshifts from literature. 
For selecting the sample of galaxies to be modeled by \galfit,
we used \sx with a background grid size BACK\_SIZE{\,=\,}64 pix
and FILTERSIZE{\,=\,}3, which we found gave enough leverage to model the background
variation without removing much light from the galaxies we wished to detect 
(the \galf modeling was done on the image without this background modeling).
The central cD in A1689 is the brightest and most extended galaxy in the cluster,
contributing to the local background of many of the
other galaxies in the FOV. As a first step in modeling its light distribution, we
masked all point and extended sources (except the cD itself) and modeled the
galaxy as a single \ser\ profile; this model was then subtracted from
the original image. A second iteration of the central galaxy model is done later,
after all the other galaxies have been subtracted (see Sec.\,\ref{ICLmodeling.sect}).

The sample of 180 galaxies (including the cD) to be fitted was selected based on size and apparent
magnitude, without considering cluster membership.  By matching these objects against
NED and the sample of A1689 galaxies studied by Halkola\etal(2006; hereafter H06), we
found that 94 of the 180 are confirmed (or very likely) cluster members based on
spectroscopic or photometric redshifts.  For each galaxy in the full sample, a region 20
times larger than the equivalent $R_e$ from \sx was extracted, and we used \galf to fit
simultaneously all the galaxies that fall in the subimage down to 1 mag fainter than the
object itself; fainter galaxies and point sources were masked. 
Due to the very high galactic density in A1689, it is common
to have contamination from neighbor galaxies that are close to, but not within, the subimage;
thus, we included the sky as a free parameter with an initial guess obtained from a
smooth large-scale background map (from a second \sx pass with a coarser background grid,
BACK\_SIZE=1024).  
We used a weight image that included instrumental and photometric contributions,
performed the fits with PSF convolution, and constrained the possible range of fitted
parameters.  We inspected the results of all fits; in some cases, we were able to
improve the fit by increasing the subimage size or changing the parameter constraints.

\begin{figure}%[!h]
\centering
\includegraphics[angle=0,width=0.44\textwidth]{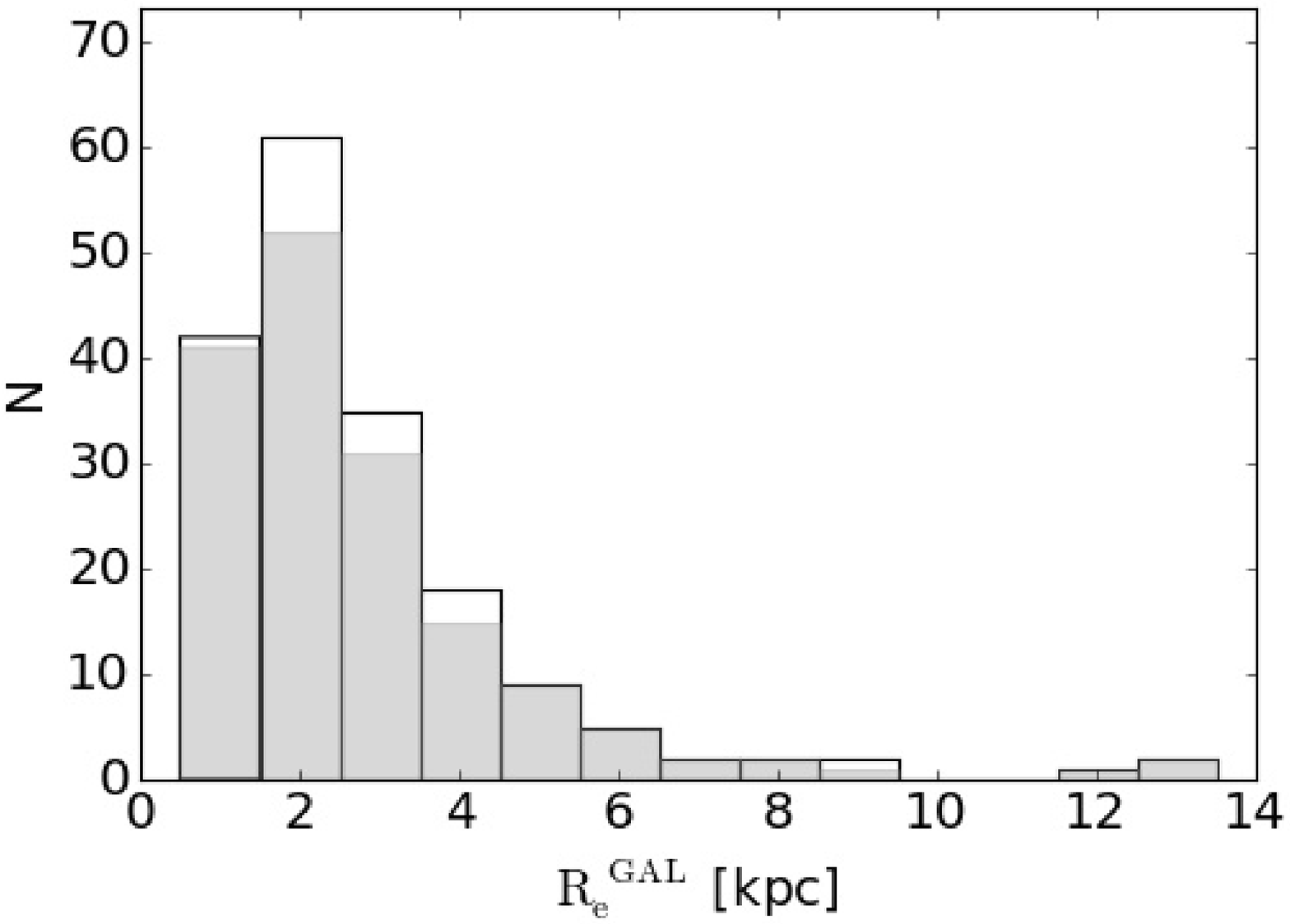}
\includegraphics[angle=0,width=0.44\textwidth]{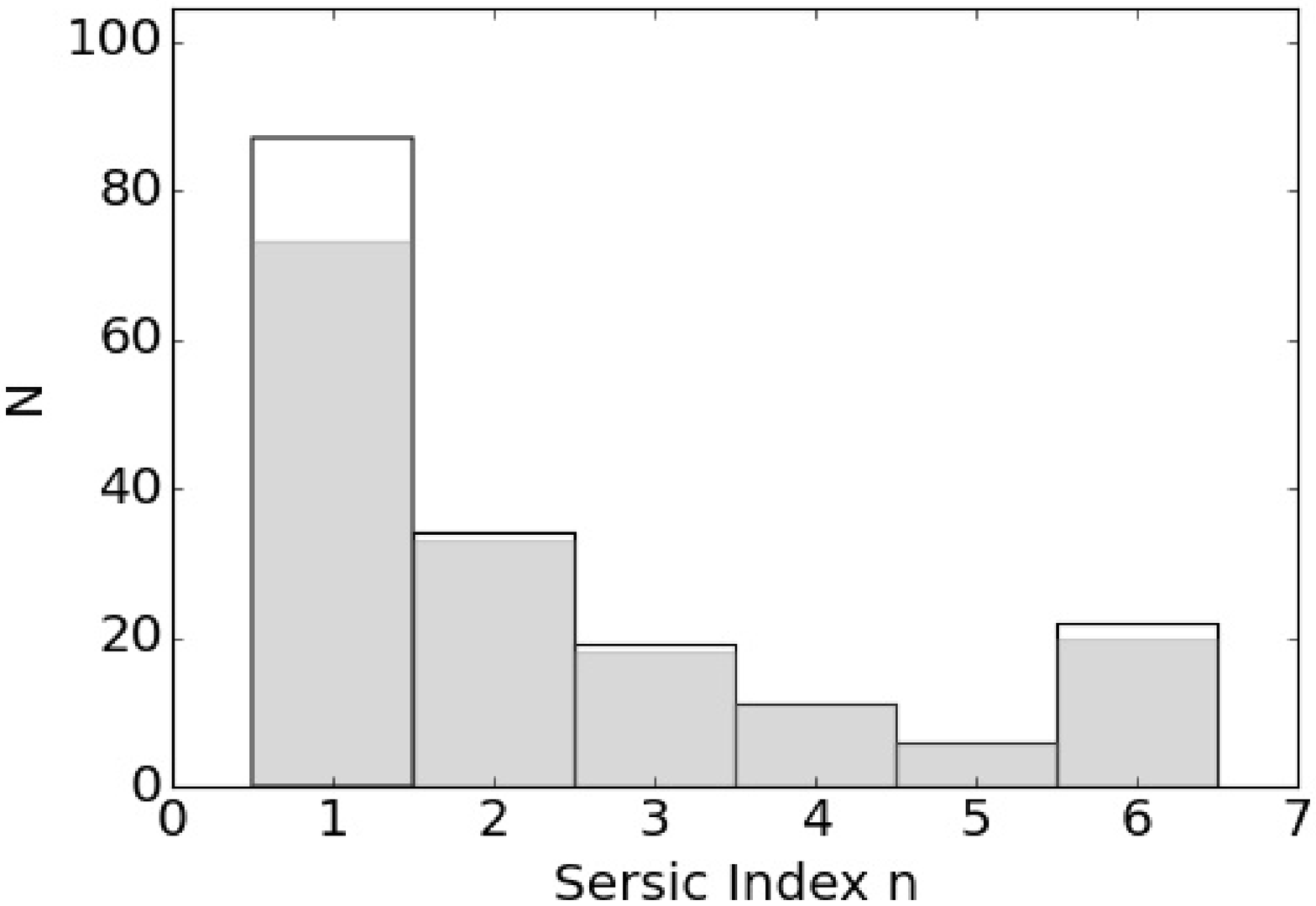}
\caption{{Top}: Histogram of $R_e^{\rm GAL}$ values for all 180 galaxies
  that were modeled in the ACS FOV (open histogram with thick black lines);
the filled gray region indicates the 
 $R_e^{\rm GAL}$ values of the 161 galaxies with fits classified as reliable.
  {Bottom}: Same as the top panel, but for the \sersic\ index~$n$. 
\label{reGal_histogram}}
\end{figure}

We performed two sets of fits: we modeled each galaxy as both a single and a double
component \ser profile.  At the very deep surface brightness levels reached by our data, 
complex structure (cores, disks, arms, rings, and tidal features) is discernible in many of the
galaxies. Because of the greater flexibility when fitting two \ser components
instead of one, the results were generally better with the double \sersic\ model.
Even for apparently smooth and regular ellipticals, the single \ser model left
notable residuals; some examples are shown in Figure~\ref{sersicFits_1c_vs_2c}.
Thus, we opted to use the results from our two-component \ser models for all the galaxies. 
In order to have a single set of parameters characterizing each galaxy, 
we take the total luminosity to be the sum of the two components and $R_e^{\rm GAL}$ 
to be the luminosity-weighted average $R_e$.
Using the robust biweight indicator (Beers et al.\ 1990), the scatter in the
differences of the total magnitudes given by the single and double \sersic\ fits for all
galaxies is 0.20~mag; the scatter estimated from the median absolute deviation
(1.48${\,\times\,}$MAD) is 0.16~mag.  Dividing these number by $\sqrt{2}$ provides an
estimate of the systematic error from the assumed model (probably the dominant uncertainty).
%% we don't really do anything with e or PA, do we?
%  and when discussing the ellipticity ($\epsilon^{GAL}$) and positional angle
%  ($PA^{GAL}$), we take the values for the more extended component.
We inspected the residuals after subtracting the double \sersic\ \galfit\ model for
each galaxy, and classified the results as
reliable (161 galaxies) or unreliable (18 galaxies).  In Figure~\ref{reGal_histogram},
we present histograms of the recovered $R_e$ and $n$ values for our fits; the
distributions are fairly typical of those found from \ser modeling
of other cluster galaxy samples (e.g., Blakeslee et al.\ 2006; Ferrarese et al.\ 2006).
The fitted photometric parameters are tabulated in Appendix~\ref{appendix:a}.

%fig2

%fig3
\begin{figure}%[!h]
\centering
\includegraphics[angle=0,width=0.46\textwidth]{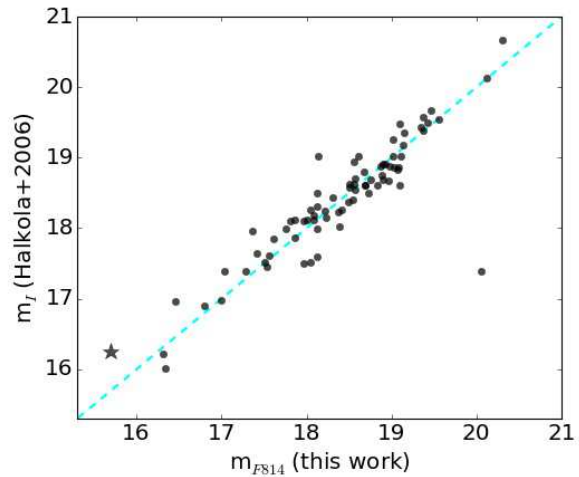}
\caption{Comparison of the total apparent F814W (${\sim\,}I$) magnitudes measured in 
  this study to the total F775W ($\sim$ SDSS $i$) magnitudes reported by
  Halkola\etal(2006). The published F775W magnitudes
  have been shifted by 0.14~mag to account for both the
  Galactic extinction and the calculated passband offset for early-type galaxies at
  $z=0.18$.  The blue dashed line indicates equality.   The star represents the cD galaxy which is model using a different approach. The galaxy with the large
  $\sim2.5$~mag offset appears to be modeled accurately in our analysis.  See text
  for details.
\label{mags_kam_halkola}}
\end{figure}

A1689 has been extensively studied because of its extreme mass, high galaxy density,
numerous arcs, and relatively modest redshift. One particular study relevant to the
present work is that of H06, who analyzed earlier ACS/WFC data
% (16\% of the exposure time with the 34\% narrower F775W bandpass)
and used a combination of strong lensing and
masses estimated for individual galaxies to derive the total and
stellar mass distributions in the cluster. To estimate the galaxy masses, they derived 
structural parameters by fitting S\'ersic profiles and assumed the galaxies followed
the fundamental plane.  
The H06 imaging data had only 16\% of the exposure time of our deep F814W images and
were taken in the F775W bandpass, which is 34\% narrower. 
The signal-to-noise of our photometric data is therefore a factor of 3~greater.

%% median offset after applying 0.1 mag is 0.06 mag with rms = 0.25 mag.
%% but they didn't apply extinction correction of 0.044 mag, so that makes
%% median offset < 0.02 mag and mean offset 0.03 +/- 0.03 mag.
%% We have T2*W2/(T1*W1) = 75172*1541.6/(11800*1023.4) = 9.6x more signal;
%% or 3.1x higher S/N.  If mag error goes as S/N, then:
%%   our e_I814 = 0.077 mag;   H06 e_I775 = 0.023 mag; 
%%

Figure~\ref{mags_kam_halkola} compares our total \Iacs\ magnitudes to the F775W
(\iacs, similar to SDSS $i^\prime$) magnitudes reported by H06.
We have applied a 0.044~mag correction for Galactic extinction (Schlafly \& Finkbeiner
2011) to the uncorrected H06 magnitudes, as well as an offset for the expected color
$(\iacs-\Iacs) = 0.10\pm0.02$~mag, calculated for the an early-type galaxy spectrum at
$z{=}0.18$.
Of the 80 galaxies they measured, 77 unambiguously match objects in our sample.
Of the H06 objects that were not matched, one is outside our field
because the observations had slightly different orientations;
another is a highly elongated, tangentially aligned object that 
we masked as a possible background arc;
and the third may have a positional error because it is several arcseconds away from
an unmatched galaxy that we modeled.
In addition, we have used a different approach to fitting the cD (see the following
section), so we use a different symbol to represent it (a star).  %, so we do not include that in the comparison. 
For the matched objects, our magnitude measurements agree very well with H06, 
except for one very deviant point, which H06 find to be 2.5~mag brighter 
than we do.  
% find $\Iacs\approx20$ while H06 reports $i\prime\approx 17.5$.
The discrepant object is a small galaxy located within a
dense region near the cluster center; examining our residuals, we are confident our
measurement is accurate.  It is likely that the H06 model for this one galaxy was
contaminated by light from its brighter neighbors, especially the cD.
but again we emphasize the overall agreement. 
The median offset (after the above extinction and color corrections) is 
0.02~mag, and the biweight scatter in the differences is 0.24~mag, or 0.22~mag
if estimated from the MAD. Assuming that the two data sets contribute 
equally to the scatter and dividing by $\sqrt{2}$ implies a
measurement error of $\sim0.15$~mag.  Given the higher signal-to-noise of our images,
this seems a conservative estimate of the errors. 

\subsection{Modeling the cD and ICL} 
\label{ICLmodeling.sect}

BCGs occupy the nexus between large-scale structures of $\gtrsim1$~Mpc, and
the $\sim\,$kpc scales of individual galaxies.  
Their structures are intricately intertwined with that of the surrounding cluster,
including the ICL, making it a challenge to disentangle properties such as
total stellar luminosity or the size of the GC population.
The cD galaxies were first identified on photographic
plates and defined by Matthews\etal(1964) as elliptical BCGs located in the centers of
galaxy clusters and surrounded by extended low surface brightness envelopes
(many were also associated with radio sources).
In A1689, the BCG has an extended, low surface brightness profile and thus is
classified as a cD; we refer to it interchangeably by either term.
However, the definition of cD has actually become less clear with the advance of
astronomical instrumentation.
The supposed unique characteristic of cD galaxies is the extended
envelope, originally characterized as an excess at large radii with respect to a de
Vaucouleurs profile (\ser with $n{\,=\,}4$), but
such a profile might also be represented by a \sersic\ of higher~$n$.
In the literature, cD profiles have been modeled as: a double de~Vaucouleurs
profile, de Vaucouleurs + exponential, \ser + exponential, and a single \ser with
$n{>}6$ (Gonzalez\etal2005; Seigar\etal2007; Donzelli\etal2011;
Bender\etal2015). Despite the diverse representations,
there is a general consensus that a
significant portion of the extended envelope results from stripping of cluster
galaxies (Gallagher \& Ostriker 1972; V\'ilchez-G\'omez 1999; Gonzalez\etal2005).
Stellar material stripped from cluster galaxies is also one definition of the ICL;
thus, in some cases, the difference between a cD envelope and ICL
may be a matter of semantics.

Kinematic information can provide insight into the transition from BCG
to extended halo or ICL, especially if there is an offset between the
velocities of the BCG and cluster.  Based on mean velocity and dispersion,
Bender\etal(2015) showed that the envelope
of the cD NGC\,6166 does not belong dynamically to the central galaxy but to the
surrounding A2199 cluster potential.  Interestingly,
although they were able to measure distinct kinematics for
the galaxy and the envelope  (or ICL), they did not find any photometric discontinuity
between these components, and the overall surface brightness profile
was well fitted with a single \ser model.
% This confirms how challenging is to disentangle the properties of both components.
% Melanie's results
% but this is difficult, especially at large distances, and/or in crowded regions
% with overlapping galaxies; A1689 is an extreme case
More generally, Veale et~al.\ (2017b) analyzed the wide-field kinematics of
a complete mass-selected sample of luminous early-type galaxies and found that 
those in cluster-sized halos preferentially show rising dispersion profiles at
large radius, while equally massive galaxies with similar profiles, but 
in more isolated regions, generally show radially declining dispersions.
Thus, disentangling a BCG halo from ICL may require
both photometric and kinematical profiles.  However, this becomes difficult
at large distances and in very dense regions with numerous overlapping galaxies,
such as in A1689.

\begin{figure}%[!h]
\centering
\includegraphics[angle=0,width=0.4\textwidth]{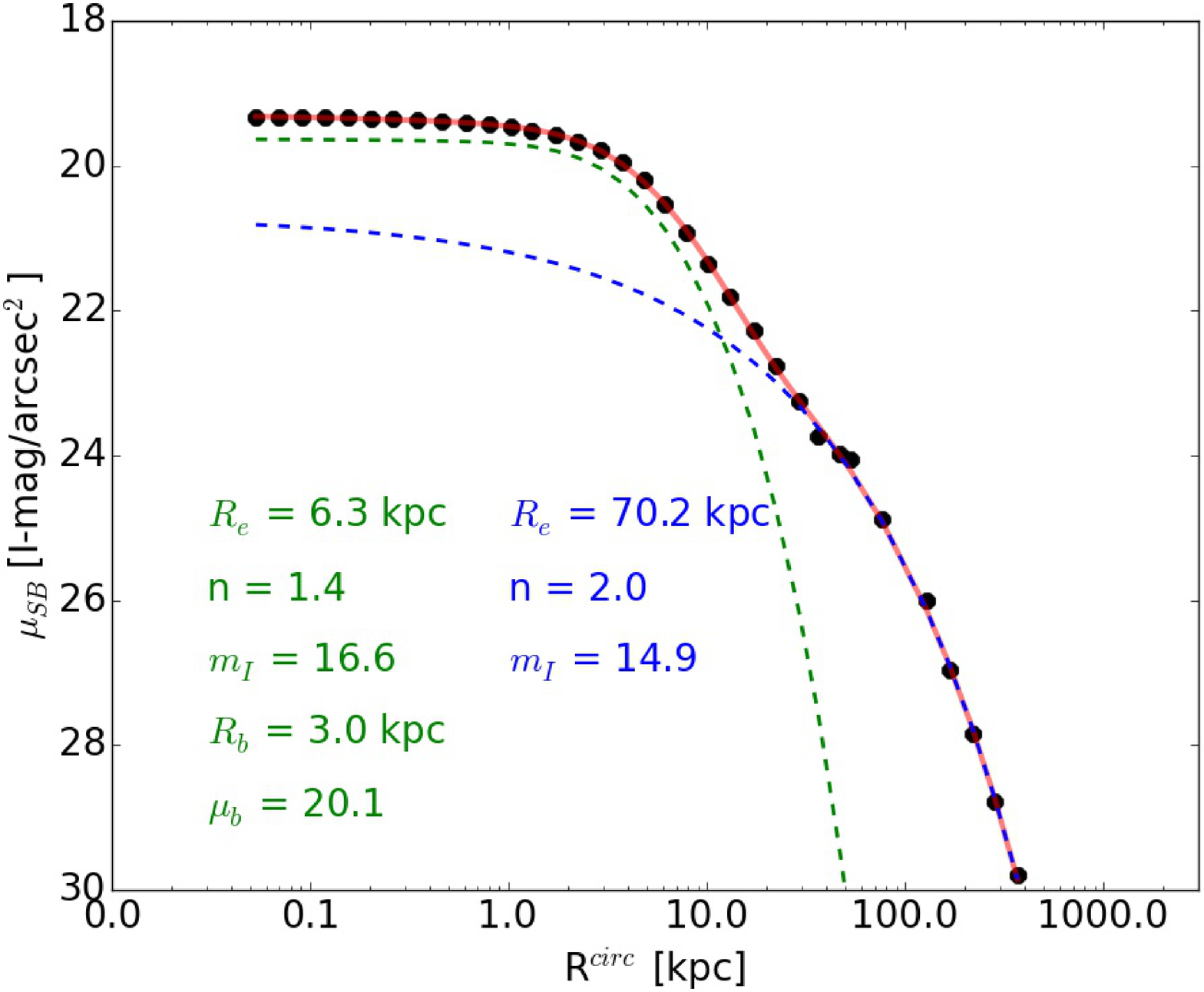}
\includegraphics[angle=0,width=0.4\textwidth]{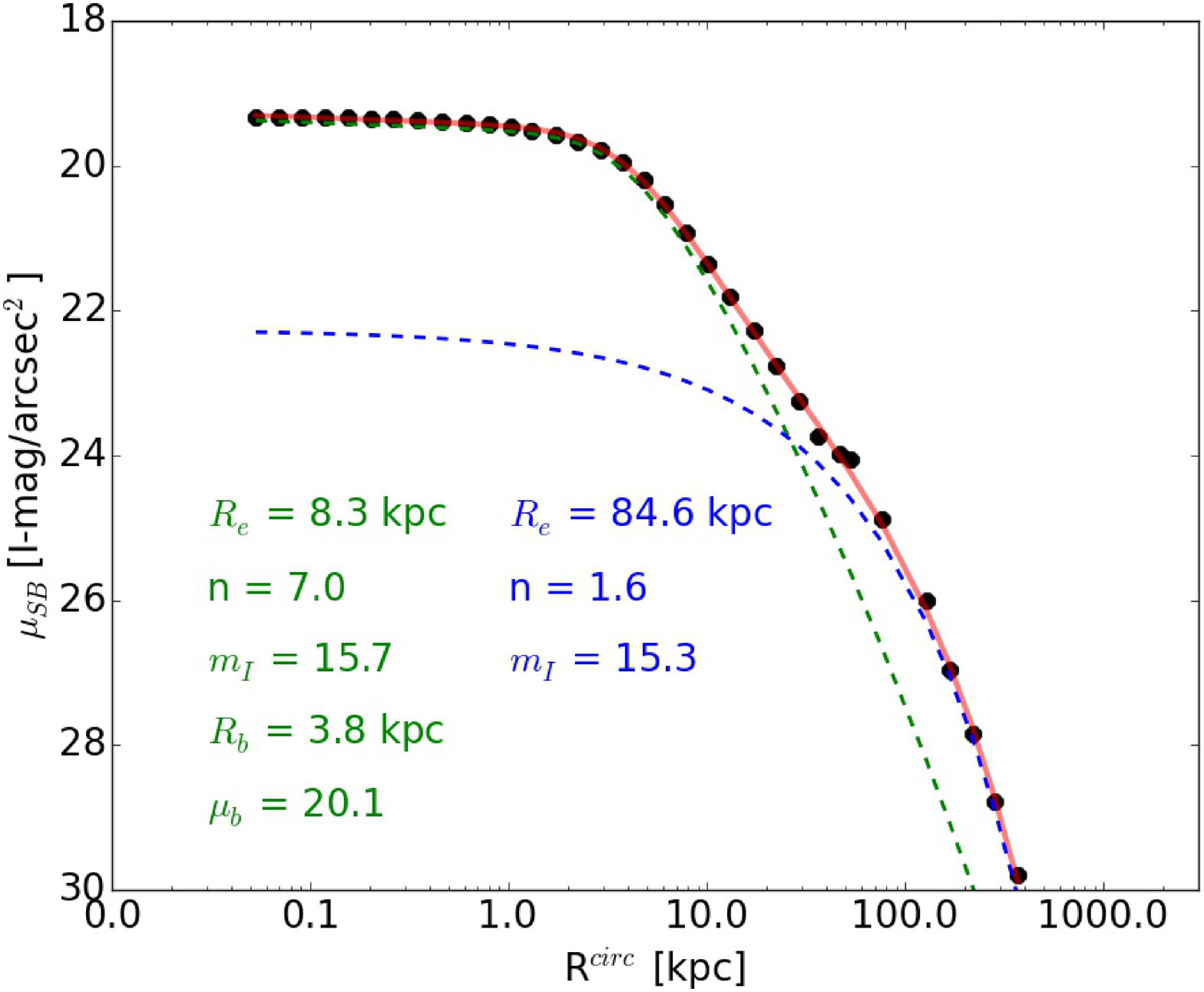}
\includegraphics[angle=0,width=0.4\textwidth]{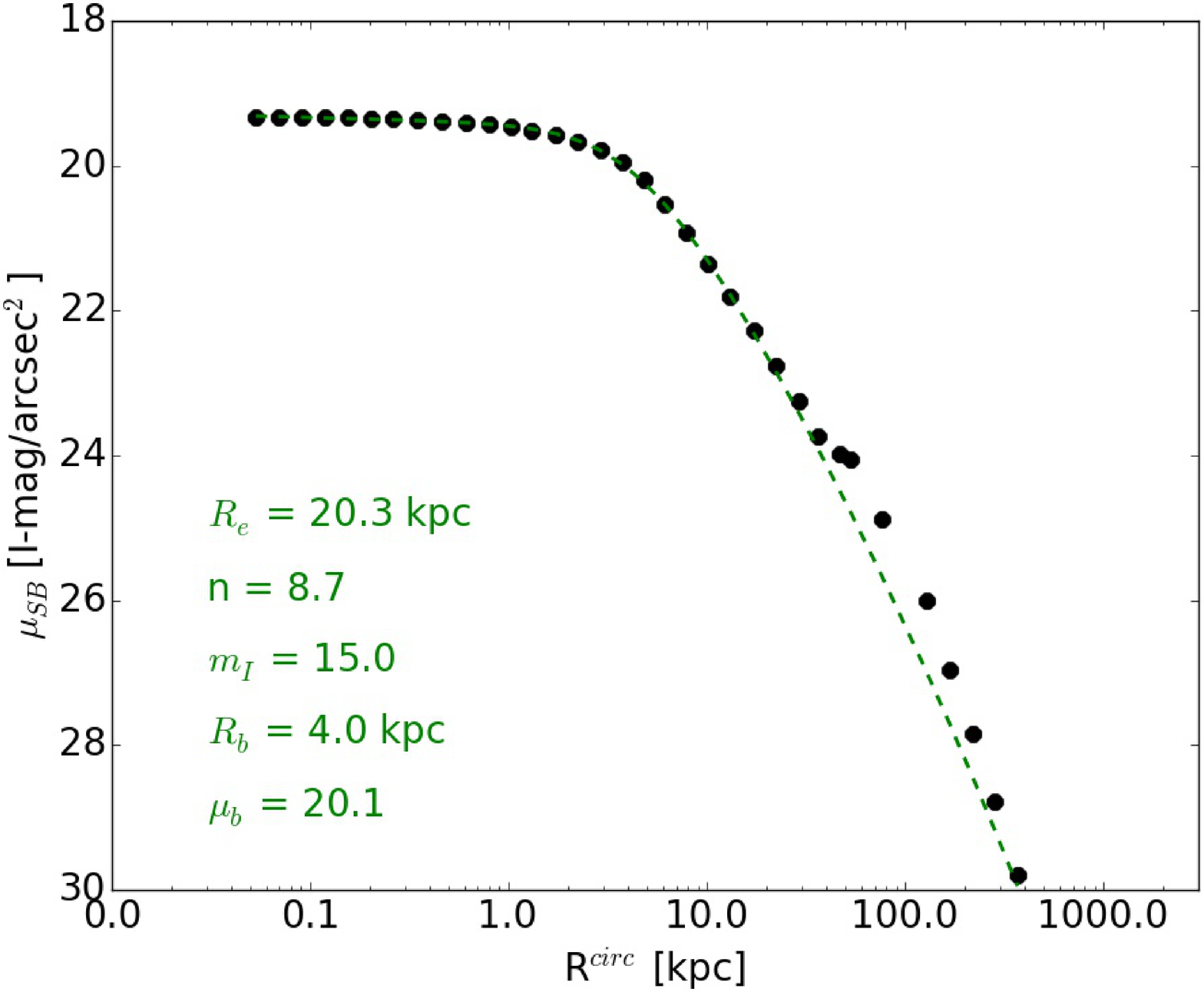}
\caption{F814W surface brightness radial profile (black dots) after
  subtracting 179 modeled galaxies in the field and masking all other sources
  except the BCG and ICL.
  In all panels, the surface brightness of each isophote is plotted as a function of
  the isophote's circularized radius.
  In the top two panels, the green dashed curve shows the best \csersic\
  fit (representing the BCG), the blue dashed curve shows the best \sersic\ fit
  (representing the ICL), and the red curve is the sum of the two.
  The bottom panel shows the result for a single \csersic\ model, fitted only to the
  isophotes with $R^{\rm circ}<25$~kpc.  The fit parameters are shown in each panel.
  The preferred decomposition model is the one in the middle panel with the $n=7$
  for the BCG; see text for details.
\label{coreSersicPlot}}
\end{figure}

In order to model accurately the extended profile of the central galaxy and extended
light, we subtracted the models for the other 179 galaxies that we had fitted, masked
other bright objects and areas of poor residuals, and used the task {\it ellipse}
within IRAF to measure the isophotes of the remaining light.  Because the galaxy is not
circular but elliptical, we calculate the equivalent circularized radius as
$R^{circ}=a\sqrt{1-\epsilon}$ where $a$ is the semi-major axis and $\epsilon$ is the
ellipticity of the isophote. In Figure~\ref{coreSersicPlot}, the black points show the
radial surface brightness profile of the central component, which includes the light
from the BCG and an obvious extra component at large radii that we attribute to ICL.~

The radial surface brightness profiles of the most massive elliptical galaxies
generally flatten at small radii, so that their central surface brightness is lower than 
the inward extrapolation of the best-fit outer \ser model.
Such profiles can be well described by a ``\csersic'' model (Graham\etal2003;
Trujillo\etal 2004), a \sersic\ profile that flattens to an inner power-law;
it has the following form:
\vbox{%
\begin{eqnarray}
I(r) \,&=& \;I_b\,2^{-({\gamma}/{\alpha})} 
\exp{\left[b_n\left(2^{1/{\alpha}}\frac{R_b}{R_{e}}\right)^{1/n} \,\right]} \\
\nonumber
&\times& {\left[ 1+\left(\frac{R_b}{r}\right)^{\alpha}\right]}^{\gamma/\alpha}
% \times \,
\exp\left\{ -b_n \left[ \frac{R^{\alpha} + {R_b}^{\alpha} }{{R_e}^{\alpha}} \right]
  ^{1/(\alpha{n})}\right\},\nonumber
\end{eqnarray} 
}
\noindent
where $R_b$ is the break radius (transition point between a standard
\ser profile and the inner power-law),
$I_b$ is the intensity at $R_b$, $\alpha$ indicates how sharp
the transition is, and $\gamma$ is the slope of the inner component.  

We tried multiple approaches for simultaneously fitting the central galaxy and ICL
with a combination of \csersic\ and \sersic\ functions, using a maximum likelihood method.
First, we performed the analysis with all
the parameters free; the resulting fit is shown in the top panel
of Figure~\ref{coreSersicPlot}.  The fit appears excellent and gives a combined
integrated magnitude of $m_{814} = 14.7$~mag; the
central \csersic\ model, representing the BCG, contributes only $\sim18$\% of the light
in this model, and the remaining $\sim82$\% is ICL. 
However, the best-fit $n=1.4$ for the BCG is lower
than expected for such a bright galaxy, while the ICL component has $n$=2.

Donzelli\etal(2011) analyzed the surface brightness profiles of
430 BCGs, finding that half of the sample required two \ser components: an inner one with
$1\lesssim n \lesssim7$, plus an outer exponential ($n\approx1$) component.
Thus, it is somewhat unusual for the ICL component to have a larger \sersic\ index
than the BCG.  Moreover, \csersic\ galaxies generally have $n>4$, even when they 
are not BCGs.  For instance, the galaxy sample analyzed by Kormendy et al.\ (2009)
included 10 classed as ``core galaxies,'' which had best-fit \sersic\ indices ranging
from 5.2 to 11.8. The largest $n$ was for M87, which might be ``biased'' by an ICL
component.  However, with M87 excluded, the remaining 9 core galaxies had a median $n$
of 7.1, with no significant correlation with luminosity (all the cored
galaxies were luminous).  This is consistent with the results of Ferrarese et al.\
(2006), who fitted multi-band profiles to 100 galaxies in the Virgo cluster, including
\csersic\ models for the brightest ones. Thus, $n\approx7$ appears to be a good average
for galaxies that are well described by \csersic\ models.

We therefore performed another fit to the combined BCG$\,+\,$ICL profile in A1689,
constraining $n=7$ for the \csersic\ component; the rest of the parameters were
left free.  The result is shown in the middle panel of Figure~\ref{coreSersicPlot}.
The combined magnitude is again  $m_{814} = 14.7$~mag (the total luminosity 
is well constrained), but in this case, the contribution from the BCG is 41\%, 
more than double the amount when all parameters were free. 
For the ICL component, we find $n=1.6$, which is very similar to the results
of Cooper et al.\ (2015), who fitted double \sersic\ profiles to high-resolution N-body
simulations of galaxy clusters and found a median best-fit value of 
$n=1.66$ for the diffuse outer component.

In order to explore the maximum amount of light that could be assigned to the
BCG, we tried a third case in which we only fitted a single \csersic\ model
to the light within $R^{circ} < 25$ kpc, and then took the integral of this fit
as the total BCG luminosity. The result is show in the
bottom panel of Figure~\ref{coreSersicPlot}.  In this case, the BCG has
$n=8.7$, which is still within the observed range, with magnitude 
$m_{814} = 15.0$~mag; since the total magnitude of the combined distribution 
is 14.7~mag, the BCG represents $\sim\,$77\% of the light in this case,
and the ICL (representing the excess of light
with respect to the single model) would have an integrated $m_{814} = 16.3$~mag.
This is a lower limit on the ICL, which must contribute at least some luminosity
even at projected radii $<25$~kpc.
Overall, we prefer the second model, with $n$ constrained 
for the \csersic\ component based on literature considerations, 
the BCG contributing $\sim\,$40\% of the light,
and the ICL accounting for the remaining $\sim\,$60\%.
We use this model in the following sections to estimate separately the specific
frequencies of the BCG and ICL.
However, it should be remembered that the plausible range for the ICL component 
is anywhere from $\sim\,$23\% to $\sim\,$82\%.

\begin{figure*}%[!h]
\centering
\includegraphics[angle=0,width=0.93\textwidth]{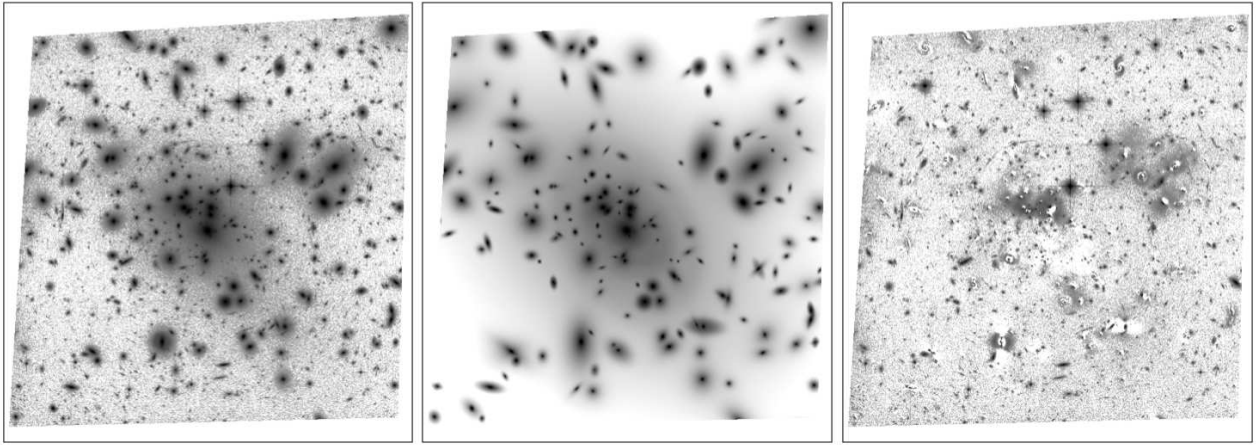}
\caption{{Left:} the deep ACS/WFC F814W image of the galaxy cluster A1689 
  (FOV$\,\sim\,$3\farcm3$\times$3\farcm3), reduced as detailed in AM13.
Middle: our final luminosity model, representing the light profiles of 180 galaxies
  and the ICL in A1689. 
Right: the image residuals after subtracting the final model from the original image.
(The residuals associated with the group of galaxies to the
  upper right of center were masked when fitting the combined light of the cD$\,+\,$ICL.)
\label{galfitModels}}
\end{figure*}

Figure~\ref{galfitModels} (middle panel) shows our final 2-D galaxy light model,
constructed from the \galfit\ models of the 179 non-central galaxies
and our {\it ellipse} model for the combined cD and ICL. The left panel of
Figure~\ref{galfitModels} shows the ACS/WFC image, and the right panel shows
the residuals after subtracting the full luminosity model.  Although only 94
of the modeled galaxies are confirmed members of A1689, the models for the
other galaxies are needed to characterize the full light distribution
and derive accurate magnitudes for the known members. Note that areas of
imperfect galaxy subtraction were masked before deriving the isophotal model
described above; thus, the excess of light associated with the clump of galaxies to
the upper right of the cD did not contribute to the ICL component seen in the 1-D
profiles of Figure~\ref{coreSersicPlot}.
% with the best fit parameters, we construct a 2D image, which is our final
% central galaxy model.  The middle panel of Figure~\ref{galfitModels} shows 
% the final model composed of 180 galaxies plus the ICL. Although only 94 are
% confirmed members of A1689, the non-member models are crucial to subtract
% contamination light from the members and derive precise total magnitudes.

%fig 4

%fig 5

%=====================   Modeling of Globular Cluster systems   ======================

\subsection{Modeling of globular cluster systems} 
\label{gcPop.sect}

 % references showing that the GCs can be well fitted by Sersic profiles 
% The GC spatial distribution is more extended than the starlight...
The situation for the GC spatial distribution is similar to the case of the starlight
in the sense that the GC populations of many galaxies in the FOV overlap.  We
therefore follow a similar approach: we use the 2-D GC number density map (described
in Sec.\,\ref{data_gc_map.sect} above) and simultaneously
fit multiple S\'ersic components  to it with \galfit.
However, we were not able to identify a significant GC system (an excess of
point sources on the map) for many of the galaxies
modeled previously from the starlight.  The main reason for this is that,
as shown by AM13, the GCs only represent $\sim0.8$\% of the stellar mass in the field,
and thus have a much lower signal-to-noise ratio than the starlight.
In addition, the Gaussian filtering, which was necessary to have a
continuous number density map, smooths out the overdensities for many
of the smaller galaxies.  The filtering was done on a physical scale of
$\sigma=13$~kpc, so only galaxies that have significant excesses
on this scale can be modeled.

\begin{figure}
\centering
\includegraphics[angle=0,width=0.4\textwidth]{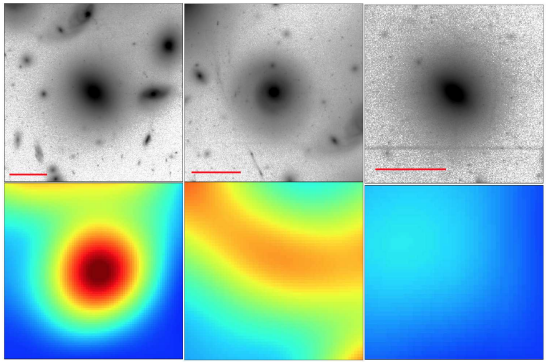}
\caption{Three examples of cases where both the galaxy and GC system were modeled.
  Top: typical modeled galaxies in the ACS field. The red bar indicates the scale corresponding to 5\,arcseconds. 
  Bottom: the corresponding regions in the GC number density map.    
The GC system of the galaxy shown in the top middle panel blends in 
with those of neighboring galaxies on either side, producing the
arc-like distribution seen in the bottom middle panel.  However, these
galaxies were simultaneously modeled in order to decompose their overlapping GC
systems into separate components. 
\label{Zoom_galaxy_GCdens.fig}}
\end{figure}

\begin{figure}
\centering
\includegraphics[angle=0,width=0.35\textwidth]{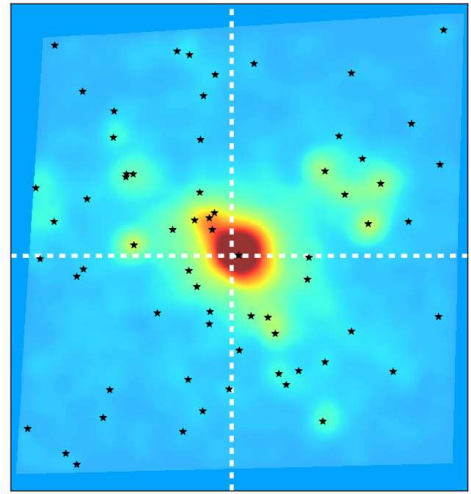}
\caption{Smoothed globular cluster surface number density map, corrected for background,
incompleteness, and masked areas. The black stars show the central
  locations of the host galaxy light for the 66 GC systems (including BCG) that were
  modeled. The dashed white lines indicate the quadrants used
  for the multiple fitting.
\label{gc_quadrants}}
\end{figure}

To select the GC systems to be fitted, we visually inspected the
subimage region of the GC number density map corresponding to each of the 180 galaxies
(including the cD) for which we had modeled the stellar light with \galfit.
We selected the cases where a significant excess was 
identifiable (see Figure \ref{Zoom_galaxy_GCdens.fig} for some
examples), resulting in a sample of 66 GC systems to be fitted.
% 65 without the cD
Because of the extensive overlapping of GC systems on the smoothed map,
rather than analyzing small
subimages around each galaxy, we divided the GC map into quadrants
(see Figure~\ref{gc_quadrants}). Within each quadrant, we performed simultaneous fitting
of all the selected GC systems within it, as well as an extended background
component for the central GC system.
Each GC system was modeled by a single \sersic\ function
convolved with the smoothing kernel used in constructing the map.
The initial guesses for the \sersic\ parameters were taken to be the best-fit
values from the galaxy models in
Sec.\,\ref{galaxymodel.sect}. The effective radius of each GC system $R_e^{\rm GC}$
was constrained to be between 0.5 and 4.0 times $R_e^{\rm GAL}$
(e.g., Peng et al.\ 2008), the central coordinates were constrained to be no more than
3\,kpc in X and Y from the galaxy center (giving a maximum offset of 4.2\,kpc, which is
large but only $\sim1/3$ of the smoothing kernel), 
and a minimum normalization of 50 GCs was imposed.
Because the GC system of the cD galaxy overlaps all of the quadrants, it was modeled
again, independently, using the full map after subtracting all other models,
as detailed below.

\begin{figure}[!h]
\centering
\includegraphics[angle=0,width=0.38\textwidth]{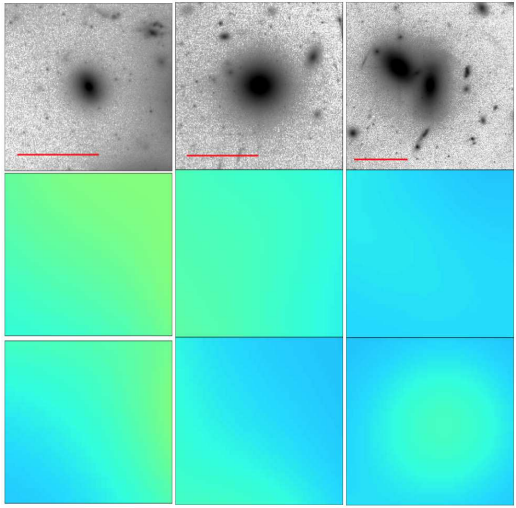}
\caption{Examples of ``bad'' GC system models, rejected on the basis
of visual inspection.
  Of the 66 GC systems modeled, 16 were labeled as bad fits,
 without considering cluster membership criteria.
Top~row: galaxies; middle row: corresponding locations on the GC density map;
bottom~row: rejected GC density model. The red bar indicates the scale corresponding to 5\,arcseconds.
\label{badGCmodels}}
\end{figure}

We inspected all 66 GC models and their residuals after subtraction from the density
map, and visually classified 50 of them as reliable fits.  Some examples of the
16~rejected fits are shown in Figure~\ref{badGCmodels}.
% good models: \ref{models_j156.fig})
Of the 50 reliable fits, 36 are associated with confirmed cluster members, 
but in three of these cases, the luminosity model was classified as poor,
leaving a sample of 33 confirmed members with good fits for both the GCs and
galaxy light.  The ratio of the normalizations of these fits 
yields the GC specific frequencies presented in Sec.\,\ref{Sn.sect}
and tabulated in Appendix~\ref{appendix:a}.

%Finally, the galaxies with reliable galaxy and GC population models are 36.  from the
% 66, 49 are 'good' fits, 36 are cluster members but there are 3 that have galaxy fit
% flag as unreliable (j=6,10,70) thus the final sample are 33 cosidering the cD

\subsection{Modeling the cD and Intracluster GCs} 
\label{gcICLmodeling.sect}

\begin{figure}%[!h]
\centering
\includegraphics[angle=0,width=0.4\textwidth]{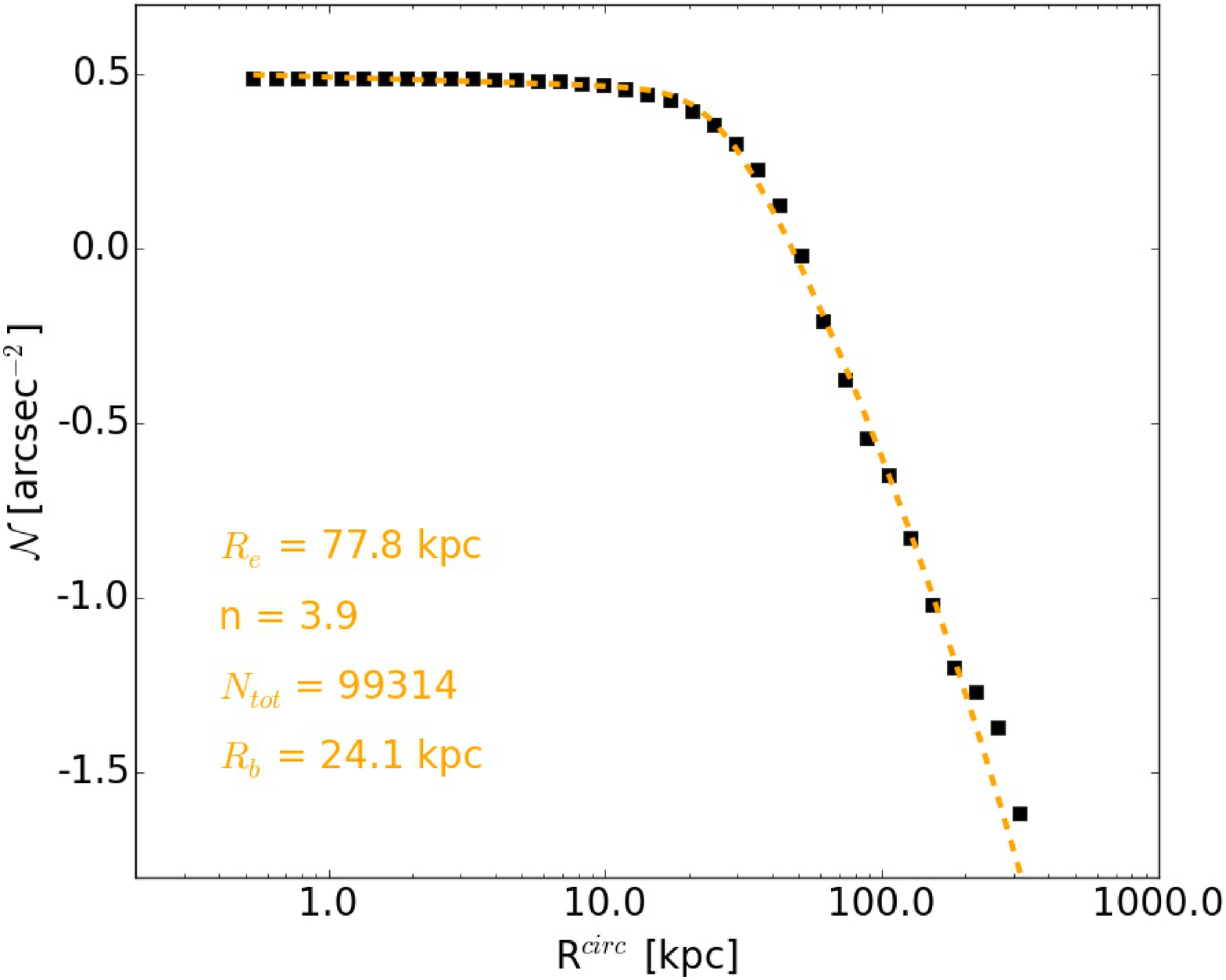}
\includegraphics[angle=0,width=0.4\textwidth]{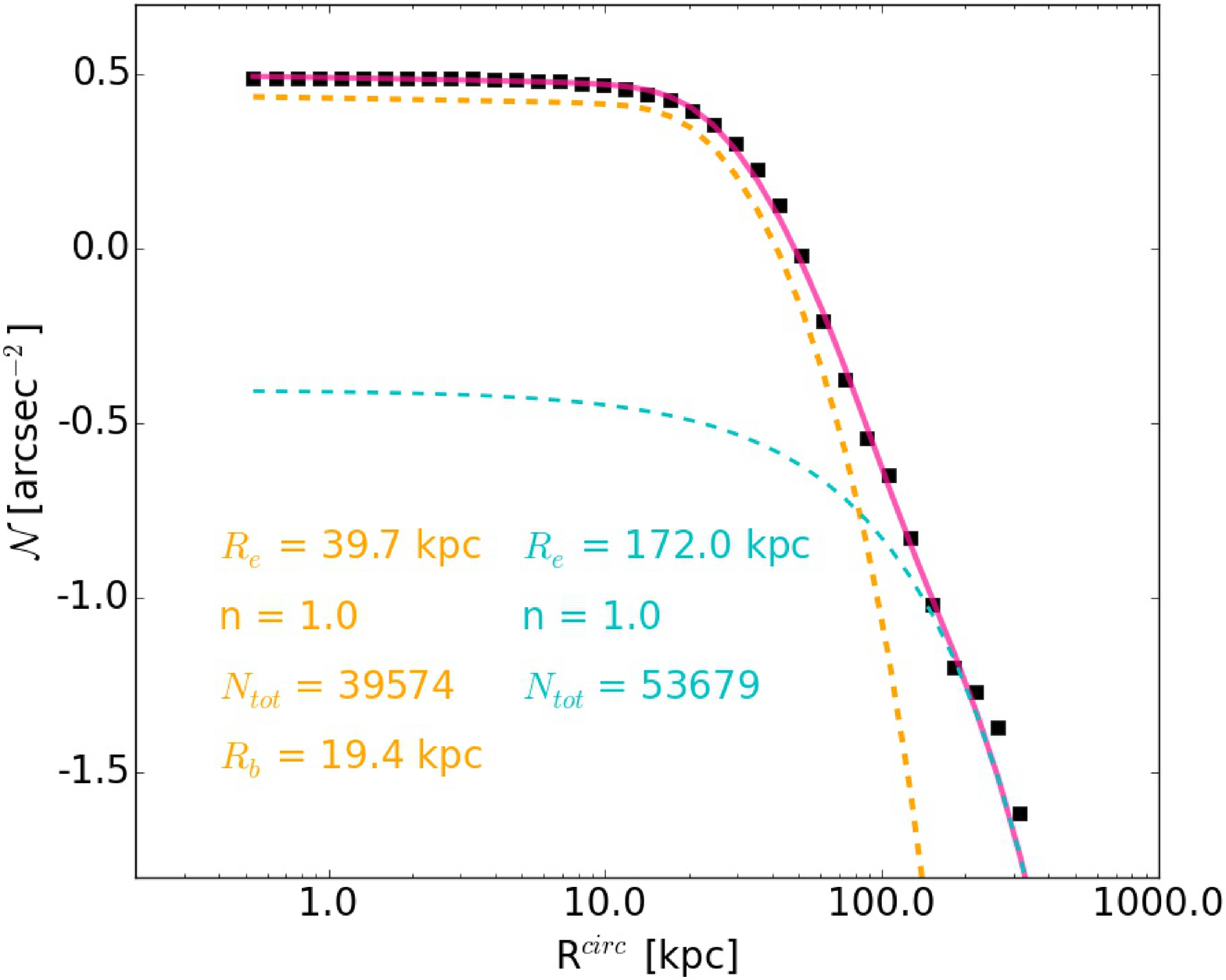}
\includegraphics[angle=0,width=0.4\textwidth]{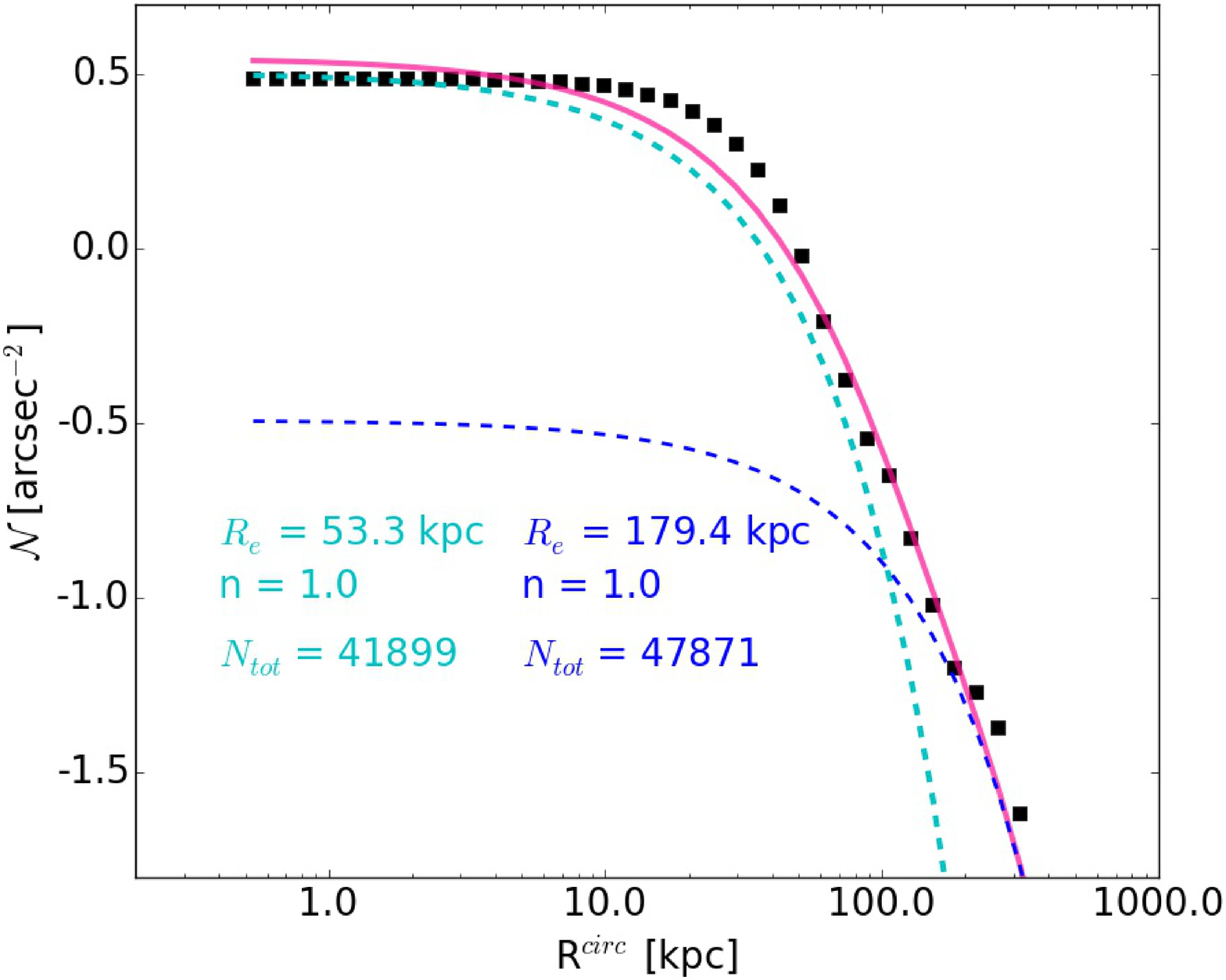}
\caption{Radial profile of the surface number density of GCs (black squares) after
  subtracting the 65 GC system models produced by Galfit and masking regions of poor
  residuals. The
  orange dashed curve shows the best \csersic\ fit (representing the GC population of
  the BCG), the cyan dashed curve shows the best \sersic\ fit (representing the ICGCs),
  and the magenta curve is the sum of the two components.  The decomposition shown in
  the middle panel is our preferred model, which
  has 42\% of the GCs assigned to the BCG component and the rest being associated
  with the ICL.
\label{gcRadialProfiles}}
\end{figure}

For the sake of consistency, we wish to follow a similar treatment for the GCs
as for the stellar light. Therefore, we
again use the {\it ellipse} task to construct the radial
profile of the surface number density of GCs for the central components (black squares
in Figure~\ref{gcRadialProfiles}, after subtracting \sersic\ models for all
the other GC systems.  We note that Durrell et al.\ (2014) and Cho et al.\ (2016)
followed a similar approach in modeling the smoothed GC number density maps
around the cD galaxies in Virgo and Coma, respectively.

\begin{figure*}
\centering
\includegraphics[angle=0,width=0.95\textwidth]{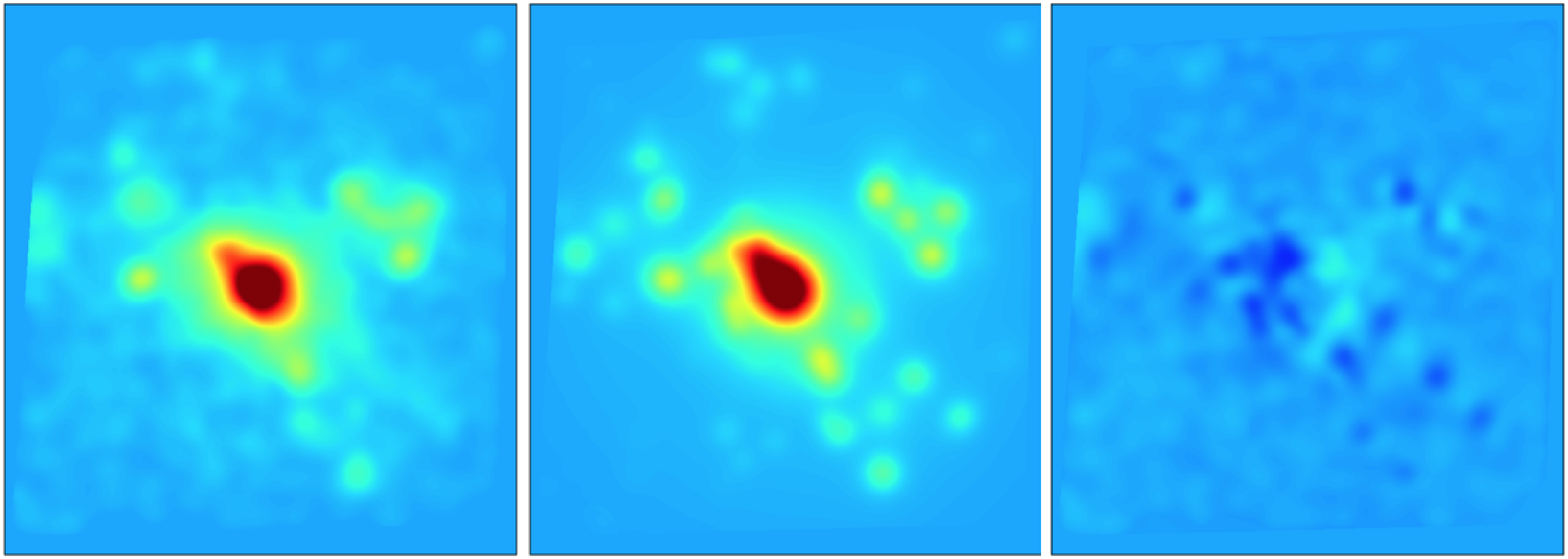}
\caption{Left: GC surface number density map (as in Fig.~\ref{gc_quadrants}).
Middle: combined models for the GC systems of individual galaxies and the ICGC component.
Right: residual after subtraction of the model.
 \label{gcDens_galfitmodel}}
\end{figure*}

Although an excess in the outer region is not as obvious for the GCs
as for the galaxy light, we again perform three types of fits,
illustrated in Figure~\ref{gcRadialProfiles}: a single
\cser (top panel), a \cser plus \ser (middle panel) and a double \ser
(i.e., testing the significance of a flat core in the GC distribution; bottom panel).
We find that the fit with two standard \sersic\ models is significantly
worse than the \csersic\ $+$ \sersic\ case; a core of $R_b \approx 20$ kpc is
required.  Somewhat surprisingly, the model prefers exponential profiles
(we constrained $n\geq1$ for both components),
whether or not there is a core.  For consistency with the light, we take the
two-component \sersic/\csersic\ model, in which $\sim42$\%
of the GCs are associated with the cD, and the other  $\sim58$\% are
ICGCs.  This is remarkably consistent with
the decomposition found in Sec.\,\ref{ICLmodeling.sect} for the central
galaxy and surrounding ICL, suggesting that the cD and ICL have
very similar specific frequencies.  However, given the similarity of the
fits in the top and middle panels of Figure~\ref{gcRadialProfiles},
the uncertainty in the decomposition is large, and a single component model
with $n\approx4$ provides a reasonable fit.
It is simply easier to separate the cD from the ICL using the starlight than the GCs,
because the surface density of the GCs is so much lower.
Figure~\ref{gcDens_galfitmodel} shows our final GC number density model 
(middle panel) compared to the observed distribution (left panel).  The residuals
(right panel) are clearly not perfect, but the model gives a reasonable representation
of the gross structure.

%===========================================================
%===========================================================
%========================   RESULTS   ========================
%===========================================================
%===========================================================

% \clearpage
\section{Results and Discussion}
\label{results.sect}

Of the 180 galaxy profiles and the 66 GC populations modeled in the FOV, 33
galaxies are confirmed A1689 members and have fits classified as reliable for both the
galaxy light and GCs. These 33, plus the intracluster components, are
discussed in the current section.

\subsection{ICL and ICGCs} 
\label{ICL.sect}

Although the ICL constitutes a significant amount of the total optical light, it is
difficult to detect because of its very diffuse and extended nature. 
Also, the ICL is centrally concentrated in the cluster, and therefore it is difficult
to distinguish from the stellar light of the central galaxy
(see Sec.\,\ref{ICLmodeling.sect}).  
There is also evidence that the ICL evolves with time in amount and morphology.  This
was noted by Schombert (1988), who found that the luminosity of the cD envelope
correlates with parameters indicative of the dynamical stage of the cluster, such as
richness, morphological type, and X-ray luminosity.
%But, hard to define... different methods, give diferent amounts for the same system
Rudick\etal(2006) studied the formation and evolution of ICL in three simulated galaxy clusters,
finding that as the galaxy cluster evolves, high luminosity features dilute into
fainter and more extended structures. 
These authors defined the ICL as all the light at $V$-band surface
brightness $\mu_{V}<26.5$ mag\,arcsec$^{-2}$, but a later study 
(Rudick\etal2011) found that the amount of ICL increased with time regardless of
definition,
although the inferred total could vary at any given time
by a factor of two, depending on one's definition.
% (equivalent to $\mu_{I}\lesssim25.8$ mag\,arcsec$^{-2}$)

In Sec.\,\ref{ICLmodeling.sect}, we found that the ICL represented
$59^{+22}_{-37}$\% of the combined light of the BCG$\,+\,$ICL in A1689, where the
substantial error bars represent the range of values from the different types of
compositions we explored.  This amounts to $11^{+4}_{-7}$\% of the total light in the
ACS FOV. This is very similar the 7\%-15\% range found by Mihos et al.\ (2017)
for the Virgo cluster, and consistent with the 10\%-40\% range found in simulations
by Contini et al.\ (2014), who also noted there was a large
halo-to-halo scatter in the ICL fraction, and no clear dependence on halo mass.

We note again that the excess in the surface brightness profile that we modeled as the
ICL (Figure~\ref{coreSersicPlot}) was not caused by the group of galaxies to the upper
right of the BCG, since we masked that region when we determined the isophotes.
%% i'm not sure if this is correct...
% In fact, the spatial distribution of the inferred ICL component (based on the ellipse
% fitting) is very round, as shown in the left panel of Figure~\ref{icl.fig}.
However, as discussed by Mihos\etal(2005, 2017), modeling the ICL 
with a simple radial profile can be misleading, since it often shows streams and other
structures that may be associated with (but not bound to) individual cluster galaxies.
Thus, although we believe we have made the best possible decomposition of A1689's
central light profile into BCG and ICL components, the amount of ICL may be
underestimated (by $\sim10$-20\%), as excess light is visible in the residual image
near the locations of some bright galaxies, and this would not be represented within
the two-component \ser model.
%

%As mentioned by Mihos\etal(2005): {\it Characterizing the ICL in
% terms of a simple radial profile will prove misleading in all but the most regular,
% relaxed clusters.} But, having a more precise value would involve a different
% analysis which is not the goal of this paper...

Similarly, although ICGCs appear to be a common feature of galaxy clusters, it is 
not easy to identify them.  The ICGCs blend in with the GC population of the BCG 
at small clustercentric radii and have low surface densities blending in with the
background at larger radii.  As noted above, the only clear detections of ICGCs have
been in relatively nearby clusters.
% of Virgo, Fornax, Coma, A1185 and Perseus.
%% if we list the clusters, also need to add all the references here.
However, numerical simulations predict that at least 30\% of the total 
GC populations in evolved clusters may be ICGCs (Yahagi \& Bekki 2005),
and this is also the fraction of ICGCs found by Peng et al.\ (2011) for the Coma
cluster.

For A1689, we have found in Sec.\,\ref{gcICLmodeling.sect} that \icgcs\ account for
about $\sim\,$58\% of the combined BCG $+$ intracluster GC population.  This is similar
to the breakdown found for the stellar light (and even more uncertain).  However,
because the $S_N$ of the central component is several times larger than average, the
\icgcs\ represent roughly 35\% of the total GC population in the ACS FOV.
AM13 did not perform any decompositions, and thus could not estimate a reliable \icgc\
fraction, but concluded that it was no more than half; our result here is consistent
with this upper limit.  In this model, the BCG contributes 26\% of the GCs to the
total, and all the other galaxies in the field contribute the remaining 39\%.

% The inferred \icgc\ population is also very round, as shown
% in the right panel of Figure~\ref{icl.fig}.

% We attribute most of the central structure in the residual images of figures
% \ref{galfitModels} and \ref{gcDens_galfitmodel} to the fact that the fit is done to
% the circularized profiles and both the light and GC distribution show some intrinsic
% ellipticity. And assuming the same center for both components (galaxy and ICL).

%% \begin{figure}%[!h]
%% \centering
%% \includegraphics[angle=0,width=0.22\textwidth]{./ICLsersic2.eps}
%% \includegraphics[angle=0,width=0.22\textwidth]{./ICGCs_from1Dcss.eps}
%% \caption{{Left:} model for the intracluster light, after subtracting the \csersic\
%%   component from the isophotal model to the residual light (black dots minus blue line
%%   in middle panel of Figure~\ref{coreSersicPlot}). {Right:} model for the ICGC
%%   component, after subtracting the \csersic\ component from the model for the residual
%%   GC number density (black dots minus blue curve in middle panel of Figure~\ref{gcRadialProfiles}).
%% \label{icl.fig}}
%% \end{figure}

%=====================   Specific frequencies   ======================
% \newpage
\subsection{Specific frequencies} 
\label{Sn.sect}

The specific frequency is defined as the number of GCs normalized to a galaxy luminosity
of $M_V=-15$ (Harris \& van den Bergh 1981):
\begin{equation}
S_N = \ngc \,10^{0.4(M_V + 15)} \,.
\end{equation}
\noindent
Early studies showed $S_N$ is dependent on morphological type (Harris 1991).  Since
then, $S_N$ has been extensively studied in galaxies with different luminosities,
morphologies, and environments, in order to understand why \ngc\ does
not scale in a simple linear way with the field stars, i.e., why the formation efficiency of GCs per
unit stellar mass varies among galaxies. This is an important key for unlocking the early
stages of galaxy formation and assembly. It is still not entirely clear
what drives the large scatter in $S_N$ for galaxies of the same mass.
An important motivation of the current work is to examine the behavior of
$S_N$ within the extremely dense environment of the A1689 core and to compare with
galaxies that have evolved under different conditions. The total mass of A1689 within 2\,Mpc is 1.3$\times10^{15}$\,\mo~(Sereno\etal2013). 
A particularly good comparison sample comes from the 100 early-type galaxies
studied in the ACS Virgo Cluster Survey (ACSVCS; \cote\ et al. 2004),
the largest sample of GC populations analyzed in a homogeneous way. Virgo has a total mass of 5.5$\times10^{14}$\,\mo (Durrell\etal2014). 
As part of the ACSVCS, Peng\etal(2008) confirmed the {\sf U}-shape trend of $S_N$ with
luminosity, and found that most dwarf galaxies with high $S_N$ are located within
1\,Mpc of the M87, the central galaxy in Virgo, suggesting an environmental dependence
of GC formation efficiency with density.  However, the innermost few dwarfs have lower
$S_N$ values, probably as a consequence of their GCs being stripped by the
strong gravitational tides of M87 or by the cluster itself (and becoming ICGCs).

% the behaviour of $S_N$ is dominated by the blue GCs

\begin{figure}%[!h]
\centering
\includegraphics[angle=0,width=0.5\textwidth]{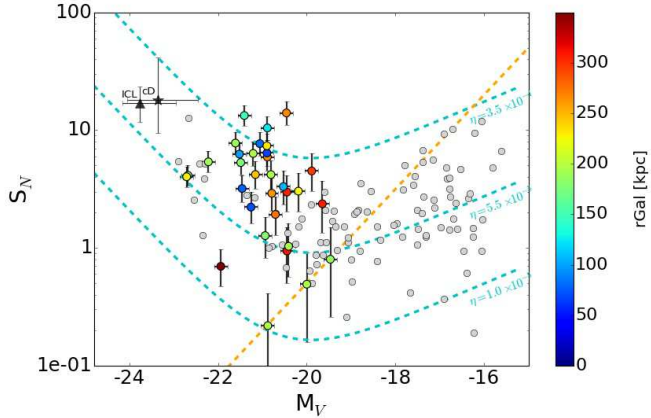}
\caption{$S_N$ as function of $M_V$ for galaxies in A1689 (colored dots, with color
  code indicating clustercentric distance according to scale at right) compared to
  same data for 100 galaxies from the ACS Virgo Cluster Survey (gray dots;
from Peng et al.\ 2008).
  The A1689 cD is represented by the labeled star, and an estimate of the intracluster $S_N$ is
  also shown. The cyan curves are predictions from Georgiev\etal(2010) assuming fixed GC
formation efficiency $\eta$ per total halo mass $M_h$ (see explanation in text);
the values of $\eta$ are labeled. The orange dashed line
indicates the lower limit of 50~GCs imposed for the A1689 GC system fits.
 \label{SN_ACSVCS_kam}}
\end{figure}

As in AM13, to estimate the $S_N$ of our A1689 sample, we convert the I-band total magnitudes
(corrected for extinction) from \galf to V-band using:  
\begin{equation}
M_V = M_{814} - K_{814} + (V{-}I_{814})\,,
\end{equation}
where $K_{814}$ is the F814W K-correction of 0.11\,mag for a giant elliptical at the
redshift of A1689, and $(V{-}I_{814})=0.83$~mag is the rest-frame color on the AB system.
Figure~\ref{SN_ACSVCS_kam} shows $S_N$ as function of $M_V$ for the ACSVCS
sample (gray dots) and our A1689 sample. The color code of the A1689 points indicates the
clustercentric distance. The star and triangle indicate the values for the
cD galaxy (${S_N}^{\rm cD}\approx18$) and the ICL (${S_N}^{\rm ICL}\approx17$), respectively.
The error bars cover the range in magnitudes from the different cases considered in
Sec.\,\ref{ICLmodeling.sect}.

The orange dashed line indicates $S_N$ as function of galaxy
magnitude for a population of 50~GCs, the lower limit imposed in the
\galf fitting; only a few systems had results actually constrained by this limit.
The cyan dashed curves are
theoretical predictions from Georgiev\etal(2010), which assume a constant GC
formation efficiency $\eta = M_{\rm GC}/M_h$ per total halo mass $M_h$, where
$M_{\rm GC}$ is the total mass in GCs.
The galaxy luminosity for these models is calculated from
% obtained $M_h$/L,
$L\propto{M_h}^{5/3}$ for galaxies of stellar mass $M_{\star}<3{\times}10^{10}$\,\mo\
(star formation presumed regulated by stellar feedback),
and L$\propto{M_h}^{1/2}$ for galaxies with $M_{\star} > 3{\times}10^{10}$\,\mo\
(star formation presumed regulated by virial shocks, including AGN-induced).
These simple scalings give a reasonable approximation to the variation in $M_h/L$ with
stellar mass.  For A1689, we find that most of the galaxies, including the BCG,
fall within the range of values observed for systems in the local universe,
indicating $1\times{10}^{-5} < \eta < 3.5\times{10}^{-4}$, although the
scatter is large and we do not probe a large range in luminosity.
As evident from the point colorings, there
is a weak tendency for galaxies with higher $S_N$ to be within $\sim$150~kpc
of the cluster center.

\begin{figure}
  \centering
% 156, m = 18.6,  S_N = 14
\includegraphics[angle=0,width=0.45\textwidth]{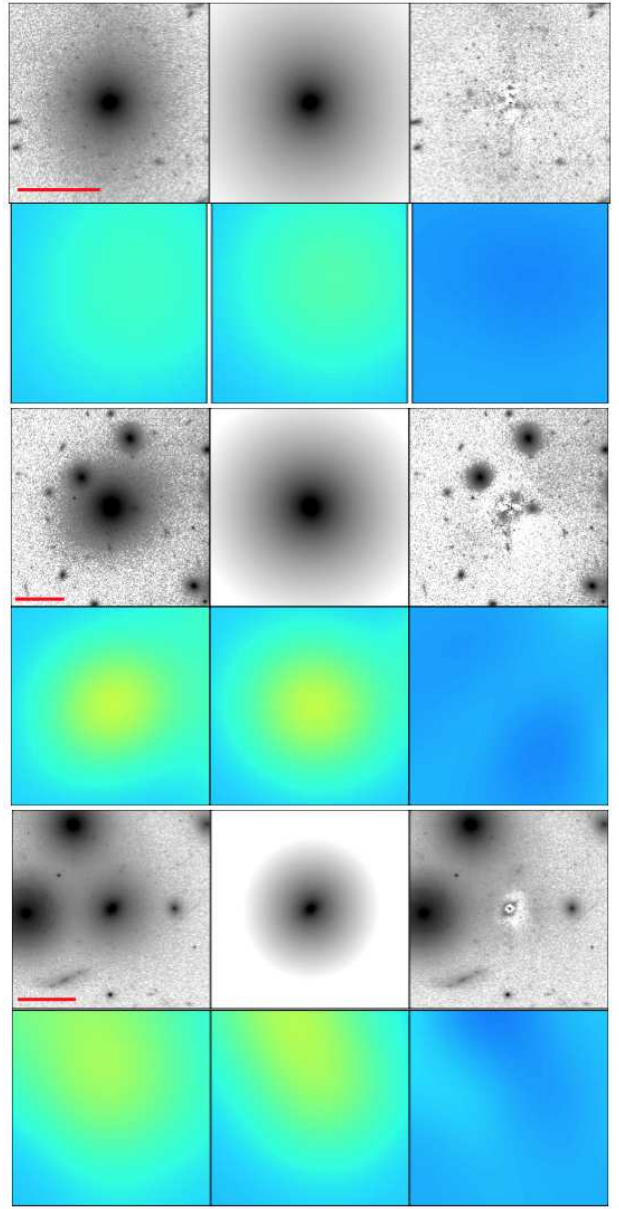}
% 63, m = 17.6, S_N = 13.3
%\includegraphics[angle=0,width=0.45\textwidth]{./fig12b.eps}
% 134, m = 18.1  S_N = 10.4
%\includegraphics[angle=0,width=0.45\textwidth]{./fig12c.eps}
%
\caption{Three intermediate luminosity galaxies with high $S_N$ values; these
  three objects are the biggest outliers with respect to the blue dashed lines
  in  Figure~\ref{SN_ACSVCS_kam}.
  Each set of six postage stamp panels shows: galaxy image, galaxy model, galaxy residual,
  GC number density map of same region; GC number density model; GC density residual. 
  The red bar indicates the scale corresponding to 5\,arcseconds. 
The galaxies are ordered from top to bottom by their $S_N$.  
The top  six panels are for galaxy \#156 with $S_N\approx14$; the middle six
are for \#63 with $S_N\approx13$; the bottom six are for \#134 with $S_N\approx10$
(the galaxy and GC fit data are listed in Tables~\ref{galaxy.tab} and~\ref{globular.tab},
respectively).  For the top two galaxies, the $S_N$ values appear genuinely
high, without obvious contamination.  For the galaxy at bottom, the GC density
distribution blends in with that of the bright galaxy to the upper left; although
both galaxies are included in the model, the decomposition may not be perfect.
The apparently bright galaxy near the left edge of the bottom panels is actually
a lower luminosity foreground spiral without detectable GCs.
\label{models_j156.fig}}
\end{figure}

The most outlying point in Figure~\ref{SN_ACSVCS_kam} is located at $M_V\approx-20.5$ and
$S_N\approx14$ and has ID~\#156 in the data tables.
This galaxy is discordant with previous results indicating that galaxies of
intermediate luminosity have universally low $S_N \lesssim2$ (Peng\etal2008).
To examine this object in more detail, the top set of 6 panels
in Figure~\ref{models_j156.fig} shows a zoomed region around this galaxy,
its corresponding location in the GC density map, the galaxy and GC models, and
residuals. Interestingly, the galaxy is $\sim250$~kpc from the cluster
center, outside the heavily crowded region.  There is no indication
that the fits are poor or misleading in this case, although the GC surface
density is fairly low.  The galaxy appears quite normal except for its
relatively high $S_N$.
%% I don't think this is true...
% However, it does appear that, while the galaxy is well-centered in the
% zoomed region, the GC population is slightly displaced, with the direction
% being towards the center of the cluster.  This could indicate
% contamination of the GC population by GCs associated with the BCG or ICL.
Figure~\ref{models_j156.fig} also shows the corresponding sets of images
for the other two outliers in  Figure~\ref{SN_ACSVCS_kam}:
galaxy \#63 (middle set of panels), which has
$M_V\approx-21.5$ and $S_N\approx13$; and galaxy \#134 (bottom panels),
which has $M_V\approx-21.0$ and $S_N\approx10$.
Only for the third object does it appear that there could be significant
contamination from a neighbor galaxy.

%=====================   D'Eugenio\etal kinematics vs. S$_N$  ======================
%\subsection{D'Eugenio\etal kinematics vs. $S_N$} 
\subsection{Galaxy kinematics vs. $S_N$} 
\label{kinematicsSn.sect}

Despite some broad morphological themes, the early-type galaxy class is
comprised of galaxies with a wide range of properties (kinematics, stellar and dust
content, mass-to-light ratio, among others).  Emsellem\etal(2007, 2011) introduced a
kinematical classification for early-type galaxies (accounting for projection
effects) by defining two kinematical classes: slow rotators (complex velocity fields with
little net angular momentum, often showing decoupled cores) 
and fast rotators (ordered velocity fields dominated by rotation).
In this classification scheme, only $\sim\,$30\% of morphologically identified
early-type galaxies are slow rotators, once considered a common 
characteristic of spheroidal systems (Cappellari\etal2011).  However, the fraction 
of slow rotators increases with mass, reaching $\gtrsim\,$80\% for stellar masses
$M_{\star}>5\times10^{11}$~\mo\ (Veale et al.\ 2017a).

\begin{figure}%[!h]
%\centering
\includegraphics[angle=0,width=0.45\textwidth]{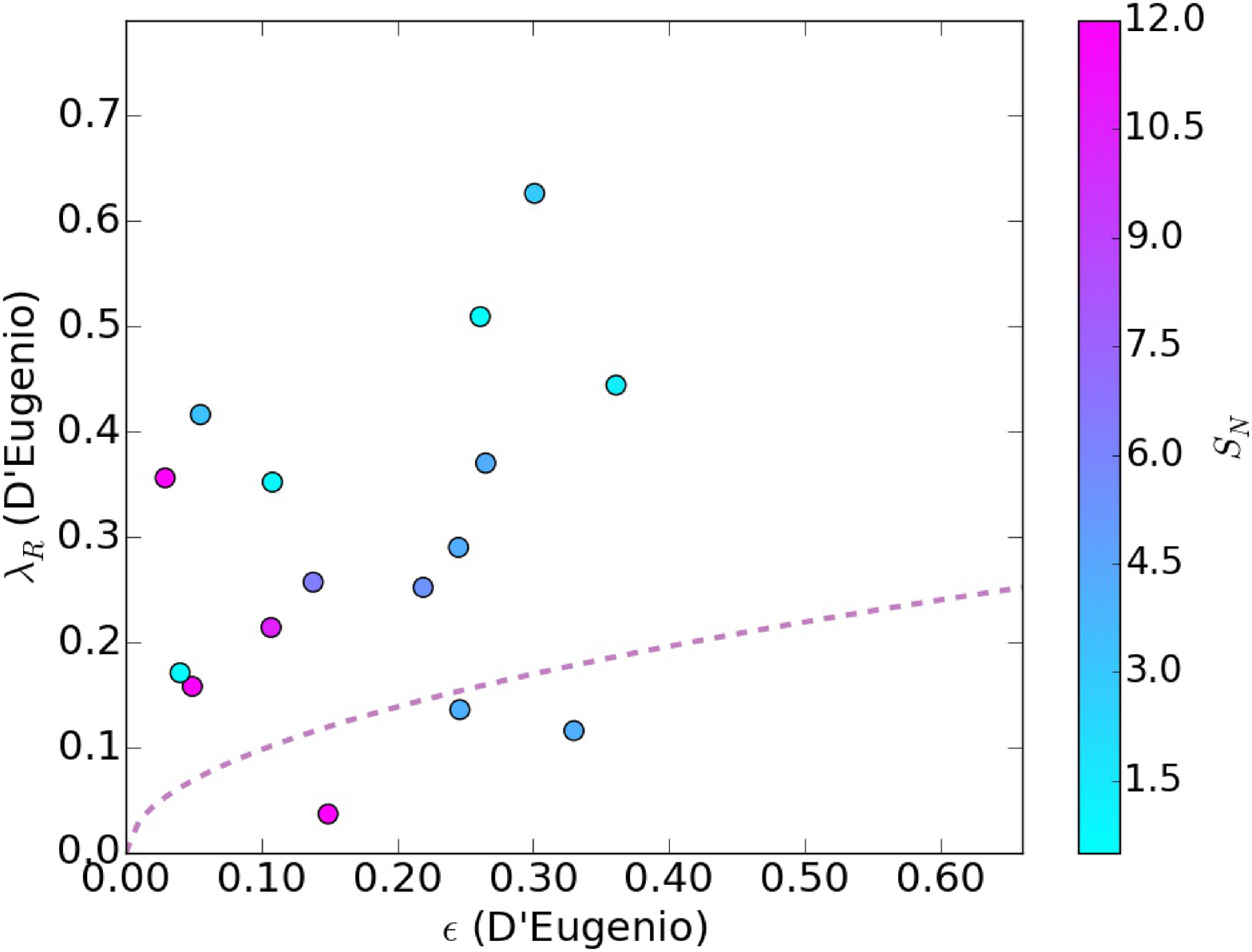}
\includegraphics[angle=0,width=0.45\textwidth]{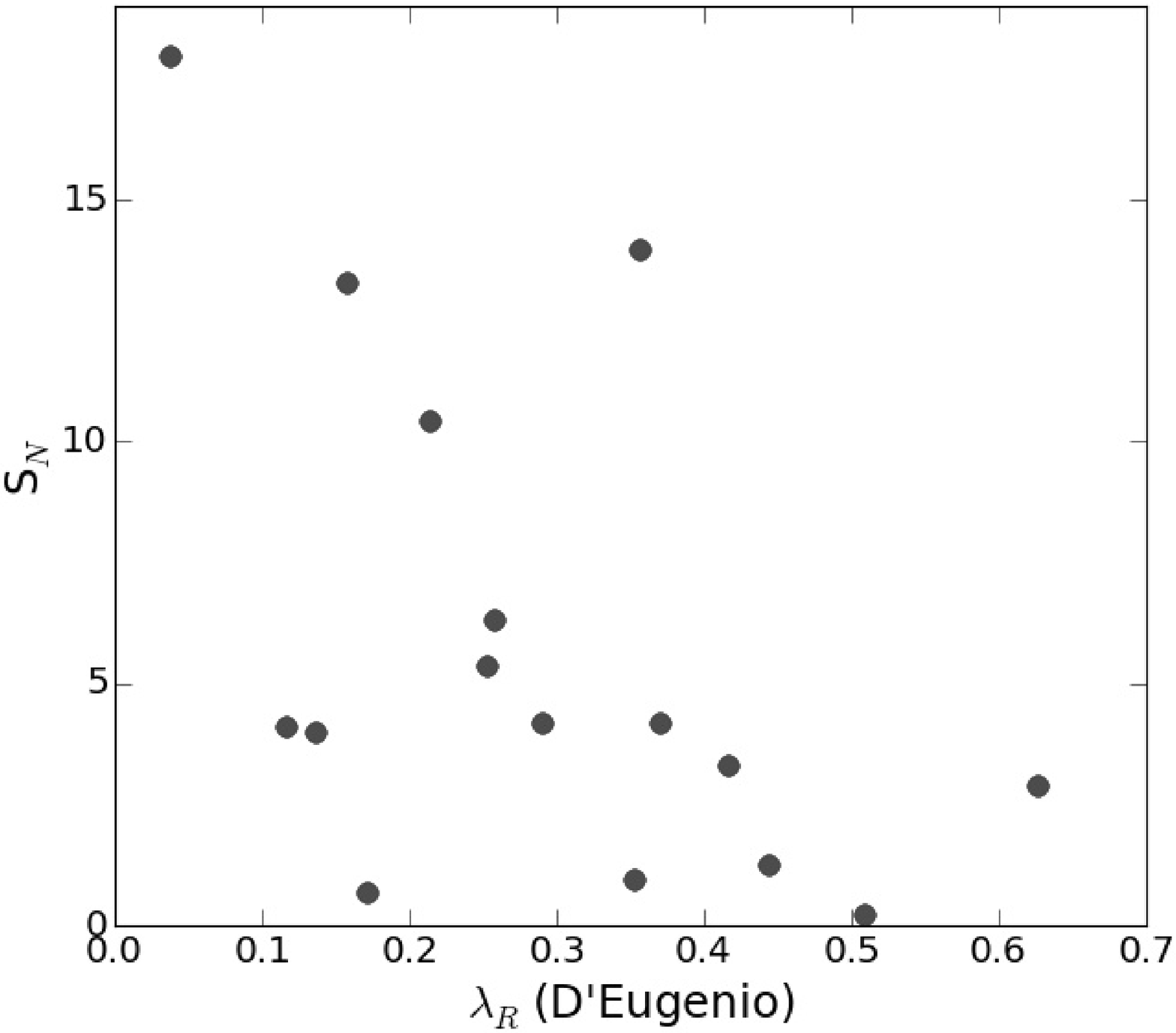}
\caption{Top: projected specific angular momentum $\lambda_R$ within $R_e$ as function of 
  galaxy ellipticity $\epsilon$ at $R_e$ (data from D'Eugenio et al.\ 2013);
the color code of the points indicate our measured $S_N$ values.  The highest $S_N$
galaxies are preferentially round.
Bottom: $S_N$ plotted versus $\lambda_R$, suggesting a possible decreasing trend;
the galaxy with $S_N=14$ and $\lambda_R\approx0.35$ is again \#156.
\label{DEugenioFig4_Sn}}
\end{figure}

The kinematical classification is based on the projected stellar angular momentum
$\lambda_{R}$ (measured within $R_e$) and projected ellipticity~$\epsilon$. The
boundary between slow and fast rotators is $\lambda_{R}=0.31\sqrt{\epsilon}$ (dashed
line in bottom panel of Figure~\ref{DEugenioFig4_Sn}), with the former being below
the line and the latter above.  This 'dichotomy' is reminiscent of the one pointed out
by Bender\etal(1989), who classified elliptical galaxies as boxy or disky based
on the deviations of their isophotes from perfect ellipses, and reported correlations
of these characteristerics with radio and X-ray emission.  Numerical simulations (Naab\etal2009;
Lagos\etal2011; Rodriguez-Gomez\etal2016) imply that dry major mergers (mass ratios of
order 3:1) form boxier, slow-rotating spheroidal systems, while wet or dry minor
mergers with high angular momentum results in disky systems. However there is not a sharp
demarcation between either the isophotal or kinematical classes, and there appears to
be a range of formation scenarios for each case.

%%%
Cappellari\etal(2011) studied the kinematical version of the morphology-density 
and found that as the galactic density increases, the fraction of early-type fast
rotators decreases in a manner similar to the decrease of spiral galaxies in the
classic morphological version of the relation (Dressler 1980).
More specifically, the fraction of slow rotators in their sample was only a few
percent, except in the Virgo cluster core, where it increased to $\sim\,$20\%.
These authors suggested that the dynamical processes operating in high-density
environments makes them more efficient in producing slow rotators.
More recently, Veale et al.\ (2017a) explored a wider range of environments and
concluded that the apparent trend was actually driven by the underlying correlation of
slow rotator fraction with stellar mass, coupled with the tendency for more massive
galaxies to occur in higher density regions.
Motivated by these considerations,
D'Eugenio\etal(2013) determined the fraction of slow rotators in the core of A1689,
a much denser environment than the Virgo cluster or any other nearby region.
These authors used FLAMES/GIRAFFE on VLT to obtain
integral field spectroscopy of 29 bright galaxies in A1689 and measure
$\lambda_{R}$, finding a similar fraction of slow rotators as in Virgo,
despite the much higher density.

Using these kinematical data, we have made a first attempt to explore the behavior of
$S_N$ as a function of $\lambda_{R}$. In the top panel of Figure~\ref{DEugenioFig4_Sn}
we reproduce Fig.\,4 of D'Eugenio\etal(2013), but only for galaxies for which
we were able to estimate $S_N$, with the $S_N$ values indicated by the color code.
Given the sample size, it is difficult to make general conclusions, but it is clear
that the slow rotators do not all have high $S_N$ values.  On the other hand, the
galaxies with high $S_N$ are all fairly round ($\epsilon<0.15$), independently of whether 
they are fast or slow rotators. Moreover, the galaxies with higher values of
$\lambda_R$ tend to have lower specific frequencies.  To highlight this latter point,
the bottom panel of Figure~\ref{DEugenioFig4_Sn} plots
$S_N$ as function of $\lambda_{R}$. The apparent anticorrelation
between these quantities is significant at the 94\%
confidence level; if galaxy \#156 (the biggest outlier in Figure~\ref{SN_ACSVCS_kam})
were excluded, the significance would increase to 97\%, although there is no \textit{a~priori}
reason to exclude this galaxy.  Such an anticorrelation between $S_N$ and $\lambda_{R}$
could occur if galaxies with low angular momentum experienced a larger number of
dissipative mergers earlier in their history and formed greater numbers of GCs in the
process, or it could indicate a lower $M_{\star}/M_h$ ratio in slow rotators, if 
\ngc\ scales reliably with halo mass.

%========================   Comparison of central coordinates of galaxy and GC system   ========================
\subsection{Comparison of centroids of galaxy and GC system} 
\label{xy_GalsGC.sect}

Numerical simulations show that when galaxies interact, their dark matter halos are
more affected than the stars (Mistani\etal2016), which are much more centrally
concentrated.  However, because the GC system is spatially more extended than the field
stars, the GCs are among the first stars to ``notice'' the interaction and undergo
stripping if enough of the dark matter halo has been disrupted (Smith et al.\ 2013).
This could be reflected as an asymmetrical shape and/or offset in the centroid of the
GCs system with respect to the starlight.   Such a displacement of the GC system 
has been found for a high-velocity early-type dwarf in the dense core of 
the Coma cluster (Cho et al.\ 2016), as well as for the cD NGC\,4874 itself.

To investigate whether the galaxies in A1689 show any evidence for offsets between the
GCs and starlight, we calculate the clustercentric radial position of each galaxy
($r_{\rm Gal}$) and its associated GC system ($r_{\rm GC}$), based on the
centroids given by the best-fit \sersic\ models for each.  We define $r$ = 0 at the
luminosity center of the cD galaxy, which coincides well with the geometrical center of
the X-ray emission.  We expect $r_{\rm Gal}$ = $r_{\rm GC}$ for undisturbed galaxies.
However, if the high-$S_N$ central GC system is augmented by stripping
GCs from cluster galaxies as they pass through the cluster core, we might expect
$r_{\rm Gal}>r_{\rm GC}$ or $r_{\rm Gal}<r_{\rm GC}$, depending on the point in the
orbit at which the galaxy is observed.  Of course, local interactions with neighboring
galaxies might also disrupt the extended GC distributions.

\begin{figure}%[!h]
\centering
\plotone{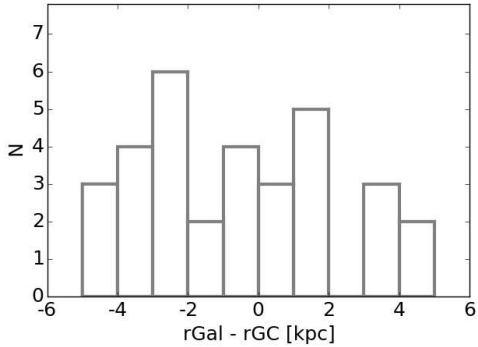}
\caption{{Top:} Histogram of differences between galaxy and GC
  system central coordinates.  Objects with
  $r_{\rm Gal}-r_{\rm GC}>0$ have GC centroids displaced in the
  direction of the cluster center; those with
  and $r_{\rm Gal}-r_{\rm GC}<0$ have GC centroids displaced away
  from the cluster center.
  The distribution appears fairly uniform; see text.
\label{dr_GalsGC}}
\end{figure}

Figure~\ref{dr_GalsGC} shows the histogram of offsets, where 
$r_{\rm Gal}-r_{\rm GC}>0$ for galaxies with the fitted GC centroid closer to
the cluster center and $r_{\rm Gal}-r_{\rm GC}<0$ for galaxies with the GC
system offset in the direction away from cluster center.  While there are more
galaxies with GC systems offset away from the cluster center, the difference is not
significant.  The distribution is quite uniform, without a strong peak at zero, likely
indicating the limitations of our GC system centroiding as a consequence of the smoothing.  
% The lower panel of Figure~\ref{dr_GalsGC} shows how the distribution of positive and
% negative GC system offsets within the ACS FOV; again we see no obvious trends.
However, we believe this is an interesting area for future studies with large samples
of GC systems in rich nearby clusters such as Coma, where it is possible to reach
farther along the GCLF, and the centroiding it not as limited by the resolution.

%
%=====================   total pop revisited   ======================
% \newpage

\subsection{Comparison to A2744 Frontier Field} 
\label{totalpop.sect}

In AM13, we found that A1689 harbored 162,850$^{+75,450}_{-51,300}$ GCs, integrated
over the GCLF, within a projected radius of 400~kpc of the cluster center.
As shown in Fig.\,3 of AM13, the ACS/WFC areal coverage is incomplete to 400~kpc;
the GCLF-integrated number within the ACS FOV is 153,600$^{+71,200}_{-48,400}$.  
By analogy with other clusters, AM13 speculated that the total GC number within the
virial radius of 3~Mpc may be several times larger, or $\sim5{\,\times\,}10^5$.
Based on these results, the A1689 GC population was the largest and most distant
that had ever been studied.  The error bars on the above values are dominated by the
uncertainty on the GCLF extrapolation, which was based on M87 (Peng et al.\ 2009)
translated to do the distances of A1689, with $K$ corrections and 2.25~Gyr of passive
evolution, and a GCLF width $\sigma=1.4{\,\pm\,}0.1$\,mag.  Harris et al.\ (2017)
advocated using $\sigma=1.3{\,\pm\,}0.1$\,mag for A1689 and similar cluster fields with
a mix of early-type galaxies; adopting this value increases the extrapolated numbers by
29\% to 198,000$^{+91,800}_{-62,400}$ GCs within the ACS FOV.

The cluster A2744 which has a total mass of $\sim$2$\times10^{15}$\,\mo \\
(Boschin\etal2006) and $z{\,=\,}0.308$, is the nearest of the Hubble Frontier Fields (Lotz et
al.\ 2017), and the brightest members of its GC system are clearly visible
(Blakeslee et al.\ 2015). Lee \& Jang (2016; hereafter LJ16) reported an extrapolated
total of 385,000~GCs, with a quoted error of only 6\%.  These authors assumed the Milky
Way GCLF with $\sigma=1.2$~mag and that the observed F814W bandpass matched
identically the rest-frame $r$-band.  They did not report the actual number of detected
GC candidates, but Harris et al.\ (2017), referencing a private communication, states
that 664 GC candidates were detected (after completeness and contamination corrections)
to $\Iacs = 29.0$~mag; this is on the Vega system and corresponds to
$\Iacs=29.4$~AB~mag, or 4~mag brighter than the GCLF turnover adopted by LJ16.
The extrapolated total based on these numbers exceeds $10^6$~GCs,
in contrast to what was reported by LJ16.
Recognizing this, Harris et al.\ used a slightly revised set of GCLF parameters for A2744 with
$\sigma=1.3$~mag and derived a total of $3.9^{+7.2}_{-2.1}\times10^5$~GCs,
where the error bars now include realistic uncertainties on the GCLF and were calculated
following the same prescription as in AM13. However, the assumed GCLF turnover magnitude
was still not fully consistent with the value adopted for the A1689 extrapolation;
using the same approach as in AM13, we estimate that the turnover should occur at an
observed AB magnitude $\Iacs = 32.9{\,\pm\,}0.2$~mag.  
For $\sigma=1.3{\,\pm\,}0.1$~mag, the extrapolation now
gives  $1.9^{+2.1}_{-0.9}\times10^5$~GCs.  With the brighter assumed turnover,
the detections go 0.4${\,\sigma}$ farther along the GCLF, lessening the extrapolation;
likewise, the error bars are smaller, in a fractional sense, than those reported by
Harris et al.\ (2017).

Given the above estimates, one can only conclude that the number of GCs in A2744 
appears similar to that found in A1689, but the value is much more poorly constrained
for A2744, with an uncertainty range at least a factor of two larger.
The fundamental reason is that AM13 detected more than ten times as many GC
candidates in A1689, and the required extrapolation was a factor ${<\,}20$, rather 
than ${>\,}200$ as in A2744.  However, there are additional reasons for caution in
interpreting the result for A2744. In the case of A1689,
the AM13 analysis accounted for the effects of Eddington bias (which causes an
overestimate of the true population), gravitational magnification of the background
counts, and passive luminosity evolution; numbers were stated explicitly for the
sizes of all these corrections.  None of these issues were even mentioned by LJ16.
Despite their quoted 6\% uncertainty, LJ16 stated that their result was
``a rough estimate of the total number of GCs in the Abell 2744 field.''
We concur on this point, and conclude that the population of GCs in A1689 remains the
largest that has been reliably measured to date.  
Future observations with the \textit{James Webb Space Telescope} (\jwst) should
reveal even larger GC systems in other high density regions.

%=====================   SMBH mass   ======================

\subsection{Core size and supermassive black hole mass} 
\label{BH.sect}

The favored explanation for the depletion of the light in the centers of massive
``core'' galaxies is the coalescence of binary supermassive black holes (SMBHs) in the
final stages of galaxy merging, and the resultant ``scouring'' of stars on orbits that
pass near the galaxy center. 
As the binary SMBHs orbit their common center of mass, stars interact
and are ejected from the system, removing angular momentum and causing the binary orbit to
shrink (Begelman\etal1980; Merritt \& Milosavljevi{\'c} 2005).
Numerical simulations of binary SMBHs predict that the
amount of depleted mass correlates with the number of mergers and the total SMBH mass
$\Mbh$ (Ebisuzaki\etal 1991; Merritt 2006).

There is an extensive body of work on
empirical relations between $\Mbh$ and properties of the host galaxy 
(or its spheroidal component),
such as luminosity, velocity dispersion ($\sigma_v$), \ser index, and \ngc\
(for reviews, see Ferrarese \& Ford 2005; McConnell \& Ma 2013; Kormendy \& Ho 2013).
Some simulations predict that the $\Mbh$-$\sigma_v$ relation gets steeper at high masses
(Boylan-Kolchin\etal2006), and this was found observationally by McConnell\etal(2011),
who reported the discovery of two SMBHs that where significantly more massive than predicted
by the $\Mbh$-$\sigma_v$ relation, suggesting a possibly different growth process at
these high masses. Rusli \etal(2013) explored the relations between depleted stellar
mass, core size, total luminosity, 
$\sigma_v$, and $\Mbh$. They found that the strongest correlation for core galaxies was
between $R_b$ (as determined from \csersic\ fitting) and $\Mbh$, having the form:
\begin{equation}
  \log\left(\frac{\Mbh}{3{\times}10^9M_{\odot}}\right) = (0.59\pm0.16) \,+\,
  (0.92\pm0.20) \log\left(\frac{R_b}{{\rm kpc}}\right) , 
\end{equation}
% 
%    10**(0.59 + .92*log10(3.8)) * 3.    -->  3.9859 x10^10
%    10**(0.59 + .92*log10(3.)) * 3.      -->  3.206x10^10
%    10**(0.59 + .92*log10(4.)) * 3.      -->  4.1784x10^10
% Thomas:  $log_{10}(\Mbh/M_{\odot})=10.27 + 1.17*log_{10}(R_b/kpc)$.
%    10**(10.27 + 1.17*log10(3.7))        -->  8.606 x10^10
%
with a quoted intrinsic scatter of 0.28~dex in mass.
From our best \csersic\ fit, we obtain a core radius $R_b=3.8\pm0.2$~kpc for the A1689
cD, which would imply an extraordinary SMBH mass $\Mbh \approx (4{\pm}2)\times10^{10}$\,\mo. 
% well into what is sometimes referred to as the ``ultramassive'' regime.
However, the above empirical scaling is not constrained for $R_b{\,>\,}1$~kpc or 
$\Mbh > 2{\times}10^{10}$\,\mo.  In fact, applying the somewhat steeper $R_b$-\Mbh\ relation from
Thomas et al.\ (2016) would imply an even more enormous \Mbh, 
greater by a factor of two, even though the relations were fitted to essentially the same data
and agree closely at $R_b<1$~kpc.  Thus, the safest conclusion is 
simply that $\Mbh$ likely exceeds $2{\,\times\,}10^{10}$\mo\ in the A1689 BCG
if its core has become depleted by the same physical process as in more nearby core
galaxies for which $\Mbh$ has been measured.

Postman et al.\ (2012) used \hst\ imaging to study the very large depleted core
in the A2261 BCG, which they described as the largest known in any galaxy.
Bonfini \& Graham (2016) fitted the same data with a \csersic\ model and determined
$R_b{\,=\,}3.6$~kpc, which they likewise characterized as the largest known. 
The latter authors suggested that depleted cores of such large sizes are
formed by mechanisms other than scouring by binary black holes, including scouring by 
a gravitationally bound group of three or more SMBHs (Kulkarni \& Loeb 2012),
a scenario that might reasonably occur in a region as dense as the center of
A1689.  In the case of the A2261 BCG, there are multiple galactic nuclei 
visible within the break radius, and Bonfini \& Graham suggest that the very large core
more likely results from the dynamical effects of these massive luminous
perturbers.  We note that McNamara et al.\ (2009) reported a core radius of
3.8~kpc in the BCG of MS0735.6+7421, although this galaxy has significant dust
in its center, and Postman et al.\ (2009) claimed the value was
overestimated by a factor of several.  For A1689, there are no other luminous
nuclei or obvious dust features within the cD's large break radius.

Another \Mbh\ correlation that has been studied in the literature is 
with the total number of GCs in the host galaxy
(Burkert \& Tremaine 2010; Rhode et al.\ 2012; Harris\etal2014). 
The most recent version has the form:
%% this is from BT2010:
% log_{10}\left(\frac{\Mbh}{M_{\odot}}\right)=(8.14\pm0.04) + (1.08 \pm
% 0.04)*log_{10}\left(\frac{N_{\rm GC}}{500}\right) 
\begin{equation}
% this is from Harris et al. 2014:
 \log\left(\frac{\Mbh}{M_{\odot}}\right)=(8.27\pm0.06) \,+\,
 (0.98\pm0.09) \log\left(\frac{N_{\rm GC}}{500}\right)   
\end{equation}
\noindent
with an intrinsic scatter of 0.30~dex in mass, after excluding galaxies
classified as S0.
For comparison, we calculate the $\Mbh$ predicted by the above relation using the GC
population of the central galaxy obtained from our preferred two-component fit
($\ngc\approx39,600$, see Figure~\ref{gcRadialProfiles}), obtaining 
$\Mbh\approx1.4^{+1.4}_{-0.7}\times10^{10}$\mo. However, if we consider the combined number
of GCs from the BCG and intracluster components ($\ngc\approx99,300$), the result is 
 $\Mbh\approx3.3^{+3.3}_{-1.7}\times10^{10}$\mo. 
Interestingly, the latter value agrees better
with the estimate from $R_b$, but as in that case, we are again exploring an unprecedented
regime requiring a large extrapolation. 
It would be extremely interesting to probe the limits of these empirical \Mbh\ correlations 
by obtaining a dynamical mass for the SMBH in A1689. 
However, given the large distance and low central surface brightness, this 
is unlikely to be possible in the near future, even with the upcoming launch of \jwst.
%\textit{James Webb Space Telescope}.

%  10**(8.14 + 1.08*log10(39574./500.))   -->  1.55x10^10
%  10**(8.14 + 1.08*log10((39574.+53679.)/500.))   -->  3.91x10^10

% connection between small and very large scales: Clarke+2017  Discovery of a giant radio galaxy in MSSS

%===========================================================
%===============   Summary & CONCLUSIONS   ================
%===========================================================

\section{Summary}
\label{final.sect}

We have modeled the light profiles of 180 galaxies near the center of A1689 using deep
\hst/ACS imaging, covering a physical area of approximately 600$\,\times\,$600~kpc at
$z{\,=\,}0.183$.  We fitted the galaxies using both single and double \sersic\ models.
Not surprisingly, the double \sersic\ fits resulted in better residuals, and we adopted
these for our final luminosity model. The photometric parameters from these fits are
tabulated in an appendix.  In order to model the luminosity profile of the central
component, including both the BCG and ICL, we first subtracted the models for the 179
other galaxies, masked all other sources, and then fitted elliptical isophotes to the
remaining light. We modeled the radial profile of the isophotes using a combination of a
\csersic\ model for the BCG and a more extended standard \sersic\ for the ICL.  Our
preferred model used $n=7$ for the \sersic\ index of the BCG and gave a break radius
$R_b=3.8\pm0.2$~kpc, at the extreme high end found for break radii in core galaxies.
The fitted \sersic\ index for the ICL in this model was $n=1.6$.  We tried a range of
fits, and concluded that the ICL represented $\sim\,$60\% of the light of the combined
profile, although in the extreme cases, it could be as little as 23\% or as much as
82\%.  This means that the ICL amounts to $11^{+4}_{-7}$\% of the total A1689 stellar
light within the FOV.

We followed a similar approach in modeling the GC systems of the galaxies, but it
proved much more challenging because GCs represent less than 1\% of the stellar light.
As a result, we were only able to obtain fits to the GC systems of 66 galaxies, and
$\sim\,$25\% of these were rejected as unreliable; we included these ``unreliable''
fits in our complete model for the GC density map, but did not use them for estimating
$S_N$ values for the galaxies in question. We omitted another $\sim\,$25\% of the
galaxies from the $S_N$ analysis because they were uncertain cluster members or had poor
luminosity fits, leaving a sample of 33 galaxies for which we have tabulated \ngc\ and
$S_N$ measurements.  Only the nearby Virgo (Peng et al.\ 2008) and Fornax (Y.~Liu, in
preparation) clusters have had $S_N$ measured for more individual galaxies.  These
A1689 galaxies are consistent with the trend of decreasing $S_N$ with increasing
luminosity found among giant ellipticals in Virgo and other nearby bright galaxy
samples.  However, a few galaxies at intermediate luminosities scatter to $S_N>10$, and
we are not able to reach lower luminosities where the $S_N$ trend reverses.

The ICGCs in A1689 are difficult to disentangle from the GCs associated with the BCG.
But, again following an analogous approach as used for the galaxy light,
we find the ICGCs comprise $\sim\,$58\% of the combined population, or about 35\% of
the GCs over the whole field.  This yields $S_N\approx18$ for both the BCG and ICL.
In this decomposition, 26\% of the GCs are associated with the BCG, and the remaining
39\% belong to non-central cluster galaxies in the FOV.
We note that we have adopted the same GCLF as in AM13, and there is an
additional uncertainty of about $\pm35$\% from the  GCLF not included in the tabulated $S_N$
measurement errors. In particular, adopting $\sigma{\,=\,}1.3$ mag as
advocated by Harris et al.\ (2017) increases our \ngc\ and $S_N$ numbers by 29\%.

Combining our results with kinematic measurements from D'Eugenio et al.\ (2013), we
find that not all ``slow rotators'' have high $S_N$.
Among the four galaxies with $S_N>10$, only the BCG/cD is a clear slow rotator.
However, all four of these galaxies are quite round, with ellipticities
$\epsilon\lesssim0.15$. 
In addition, there appears to be a trend, significant at the 94\% level,
of decreasing mean $S_N$ with increasing specific angular momentum $\lambda_R$. 
These results are consistent with the observed trend of
lower $S_N$ in S0 galaxies as compared to giant ellipticals.
An anticorrelation between $S_N$ and $\lambda_{R}$ might result if galaxies with lower
angular momenta experienced many dissipative mergers during their formation
and consequently formed GCs with greater efficiency; alternatively,
it could indicate a lower field star formation efficiency in slow
rotators, if \ngc\ is a reliable tracer of halo mass.

We also reevaluated the GC population in A2744, the nearest of the Hubble Frontier
Fields at $z{\,\approx\,}0.31$. LJ16 claimed that it had the largest known population
of GCs with a quoted uncertainty of only 6\%.  However, because of its larger distance,
fewer than one-tenth as many GCs have been detected in A2744 as compared to A1689.  
Using a consistent set of assumptions for the GCLFs in these two clusters, we find
similar extrapolated total GC numbers, but the uncertainty on the extrapolation is much
greater for A2744.  We conclude the GC population in A1689 remains the largest that has
been reliably measured.  This will be an interesting area to revisit with \textit{JWST}.

The unusually large core in the A1689 BCG may hint at a central SMBH more
massive than any yet detected by dynamical techniques in nearby galaxies.
Published $\Mbh$-$R_b$ scaling relations are not empirically constrained at 
$R_b{\,\gtrsim\,}1$~kpc, but a naive extrapolation to $R_b{\,=\,}3.8$~kpc would imply
$\Mbh\sim4{\times}10^{10}$~\mo.  We also  
calculated the SMBH mass implied by an extrapolation of the published scaling
relation between \Mbh\ and \ngc, although in this case there can be no causal connection.
Because sizable extrapolations of the scaling relations are required in both cases,
we simply conclude that $\Mbh>2{\times}10^{10}$\,\mo, if the core in this galaxy
grew through the same processes that depleted the cores in nearby galaxies with
measured SMBH masses. This is a question that likely must await high-resolution
dynamical studies with the coming generation of extremely large telescopes.

\acknowledgments
K.A.M.\ acknowledges support from FONDECYT Postdoctoral Fellowship Project No.~3150599.
We thank Bill Harris, Eric Peng, Thomas Puzia, Patrick \cote, and Laura Ferrarese for
helpful discussions.
\facilities{HST (ACS/WFC)}

\appendix

\section{Tables of Photometric Parameters and Specific Frequencies}
\label{appendix:a}

Table~\ref{galaxy.tab} tabulates the \sersic\ parameters for the 180 galaxies that we
modeled in A1689. As discussed in Sec.~\ref{galaxymodel.sect}, we estimate the 
error in total magnitude of each galaxy to be about 0.15~mag.
Table~\ref{globular.tab} lists the total number of globular cluster
\ngc, fitted \sersic\ parameters, and specific frequency for each GC system in A1689
that we analyzed; the ID numbers can be used to match these GC system measurements
with the photometric parameters for corresponding galaxies in Table~\ref{galaxy.tab}.

%\begin{longtable*}{*{10}{c}}
%\multicolumn{10}{c}  {Galaxy \ser parameters}  \\
%\noalign{\smallskip}\hline
%${\rm ID}$ &  RA & DEC & $m_{F814}$ & $R_e^{\rm GAL}$ & \ser index & PA &  $\epsilon$ & cluster  &  flag  \\
%  &    &   &   &   [kpc] &  &   &   &   member  \\
%\noalign{\smallskip}\hline
%\input{./table1.txt} 
%\noalign{\smallskip}\hline
%\label{galaxy.tab}\end{longtable*} 

% j   RA   DEC   mTotal    m1   Re1   n1  e1   m2   Re2   n2   e2   cluMember   fitFlag
\startlongtable
\begin{deluxetable}{*{14}{c}}
\tablecaption{Galaxy \ser parameters 
\label{galaxy.tab}}
\tablehead{ ${\rm ID}$ &  RA & DEC & $m_{F814}$ & $m^{\rm c1}$ & $R_e^{\rm c1}$ & n$^{\rm c1}$ &  $\epsilon^{\rm c1}$ & $m^{\rm c2}$ & $R_e^{\rm c2}$ & n$^{\rm c2}$ &  $\epsilon^{\rm c2}$ &  cluster  &  flag \\
  &    &   &   & &  [kpc] &  &   &   &  [kpc] & & &  member  &  }
%\tablehead{ ${\rm ID}$ &  RA & DEC & $m_{F814}$ & $R_e^{\rm GAL}$ & \ser index & PA &  $\epsilon$ & cluster  &  flag  \\
%  &    &   &   &   [kpc] &  &   &   &   member  }
\startdata
% j   RA   DEC   mTotal    m1   Re1   n1  e1   m2   Re2   n2   e2   cluMember   fitFlag  
   1  &   197.86360   &  -1.33608   &    17.4   &    17.6   &   5.6   &   0.7   &  0.06   &    19.7   &   1.1   &   1.2   &  0.02   &   n   &   1  \\ 
   4  &   197.85226   &  -1.34072   &    20.6   &    20.9   &   4.5   &   0.7   &  0.29   &    22.2   &   1.5   &   1.0   &  0.34   &   u   &   1  \\ 
   5  &   197.88098   &  -1.32576   &    16.8   &    16.9   &   5.1   &   1.4   &  0.21   &    19.4   &   1.0   &   0.7   &  0.14   &   y   &   1  \\ 
   6  &   197.85576   &  -1.34360   &    17.0   &    17.0   &   9.7   &   2.5   &  0.46   &    20.6   &   0.8   &   1.7   &  0.38   &   y   &   3  \\ 
   8  &   197.86307   &  -1.34953   &    18.9   &    19.4   &   4.4   &   0.5   &  0.49   &    19.9   &   1.8   &   0.7   &  0.60   &   y   &   2  \\ 
   9  &   197.88348   &  -1.34679   &    20.9   &    20.9   &   3.2   &   1.1   &  0.68   &    24.0   &   1.4   &   1.8   &  0.70   &   u   &   1  \\ 
  10  &   197.89201   &  -1.35052   &    17.9   &    18.1   &   4.3   &   1.9   &  0.60   &    19.6   &   4.0   &   4.0   &  0.60   &   y   &   3  \\ 
  12  &   197.89438   &  -1.36456   &    18.3   &    18.4   &   5.0   &   1.3   &  0.14   &    20.9   &   0.3   &   1.4   &  0.26   &   u   &   1  \\ 
  13  &   197.85151   &  -1.31046   &    19.3   &    20.2   &   4.0   &   0.5   &  0.60   &    19.9   &   2.6   &   4.0   &  0.58   &   y   &   1  \\ 
  15  &   197.87502   &  -1.34447   &    18.1   &    18.3   &   4.5   &   0.6   &  0.36   &    19.8   &   0.5   &   1.5   &  0.05   &   y   &   2  \\ 
  17  &   197.90275   &  -1.32853   &    19.1   &    19.9   &   2.4   &   5.6   &  0.40   &    19.8   &   3.1   &   0.8   &  0.20   &   y   &   2  \\ 
  18  &   197.86689   &  -1.34326   &    20.8   &    21.1   &   3.6   &   0.7   &  0.60   &    22.4   &   1.2   &   1.2   &  0.60   &   u   &   1  \\ 
  19  &   197.89844   &  -1.33672   &    18.4   &    18.5   &   4.3   &   0.7   &  0.19   &    20.7   &   0.1   &   4.0   &  0.30   &   y   &   3  \\ 
  20  &   197.86696   &  -1.32451   &    19.3   &    19.4   &   4.0   &   1.8   &  0.55   &    21.8   &   3.3   &   4.0   &  0.58   &   u   &   2  \\ 
  22  &   197.89740   &  -1.35921   &    17.7   &    18.0   &   5.6   &   1.2   &  0.60   &    19.2   &   0.8   &   1.7   &  0.30   &   y   &   2  \\ 
  23  &   197.86690   &  -1.31216   &    18.7   &    19.4   &   5.0   &   0.6   &  0.69   &    19.5   &   1.4   &   2.9   &  0.37   &   y   &   2  \\ 
  24  &   197.86216   &  -1.32797   &    19.5   &    20.9   &   5.6   &   0.5   &  0.00   &    19.9   &   3.4   &   2.1   &  0.68   &   u   &   2  \\ 
  25  &   197.86820   &  -1.31244   &    18.7   &    19.1   &   4.3   &   0.6   &  0.60   &    20.0   &   0.6   &   2.7   &  0.39   &   y   &   2  \\ 
  27  &   197.85645   &  -1.33895   &    21.0   &    22.3   &   2.0   &   1.6   &  0.35   &    21.4   &   2.2   &   0.7   &  0.15   &   u   &   1  \\ 
  28  &   197.87990   &  -1.31331   &    21.1   &    22.4   &   1.0   &   1.0   &  0.33   &    21.5   &   1.8   &   1.1   &  0.11   &   u   &   1  \\ 
  29  &   197.88623   &  -1.33292   &    16.3   &    16.8   &  10.0   &   1.5   &  0.39   &    17.6   &   1.6   &   2.4   &  0.21   &   y   &   1  \\ 
  30  &   197.88147   &  -1.32351   &    19.0   &    19.4   &   3.8   &   0.8   &  0.57   &    20.2   &   0.6   &   4.0   &  0.09   &   y   &   2  \\ 
  32  &   197.85839   &  -1.32649   &    19.6   &    21.5   &   2.5   &   0.5   &  0.21   &    19.8   &   2.7   &   2.9   &  0.07   &   y   &   1  \\ 
  33  &   197.88642   &  -1.32545   &    16.3   &    16.5   &  15.2   &   4.5   &  0.42   &    18.3   &   2.0   &   3.7   &  0.18   &   y   &   1  \\ 
  35  &   197.88095   &  -1.35931   &    20.1   &    20.7   &   7.2   &   3.2   &  0.60   &    21.1   &   1.9   &   0.5   &  0.54   &   u   &   1  \\ 
  36  &   197.87263   &  -1.34536   &    19.5   &    19.9   &   2.5   &   0.5   &  0.36   &    20.7   &   0.3   &   2.1   &  0.14   &   u   &   3  \\ 
  38  &   197.87536   &  -1.34522   &    16.4   &    16.5   &  12.9   &   6.0   &  0.16   &    19.4   &   5.1   &   0.5   &  0.10   &   y   &   2  \\ 
  39  &   197.86988   &  -1.31737   &    19.1   &    19.8   &   4.1   &   0.7   &  0.39   &    19.9   &   1.3   &   4.0   &  0.46   &   y   &   1  \\ 
  40  &   197.88324   &  -1.34428   &    19.8   &    20.1   &   4.6   &   0.8   &  0.44   &    21.3   &   0.7   &   2.0   &  0.26   &   u   &   1  \\ 
  41  &   197.87525   &  -1.34123   &    17.3   &    17.8   &   7.3   &   0.8   &  0.20   &    18.3   &   1.5   &   1.4   &  0.22   &   y   &   2  \\ 
  42  &   197.86826   &  -1.33290   &    18.0   &    18.3   &   2.9   &   4.3   &  0.44   &    19.5   &   5.1   &   0.7   &  0.58   &   y   &   2  \\ 
  43  &   197.87130   &  -1.36038   &    18.6   &    18.9   &   3.7   &   0.9   &  0.60   &    20.1   &   0.5   &   1.4   &  0.30   &   y   &   3  \\ 
  44  &   197.88038   &  -1.33774   &    19.7   &    19.8   &   3.6   &   3.5   &  0.60   &    23.2   &   0.6   &   1.1   &  0.60   &   u   &   2  \\ 
  45  &   197.87314   &  -1.35459   &    20.7   &    21.6   &   4.0   &   0.7   &  0.23   &    21.2   &   1.3   &   1.5   &  0.13   &   u   &   1  \\ 
  46  &   197.86932   &  -1.34069   &    18.4   &    18.5   &   6.1   &   3.2   &  0.48   &    21.8   &   0.3   &   1.7   &  0.18   &   y   &   2  \\ 
  47  &   197.87417   &  -1.33881   &    20.0   &    21.4   &   1.8   &   0.6   &  0.27   &    20.4   &   5.1   &   4.0   &  0.48   &   u   &   1  \\ 
  48  &   197.88292   &  -1.31491   &    18.9   &    19.2   &   2.6   &   4.8   &  0.09   &    20.7   &   3.8   &   0.6   &  0.13   &   y   &   1  \\ 
  49  &   197.84620   &  -1.35477   &    18.5   &    18.8   &   4.1   &   1.3   &  0.60   &    19.9   &   0.4   &   2.9   &  0.23   &   y   &   3  \\ 
  50  &   197.87263   &  -1.31180   &    20.1   &    21.8   &   1.3   &   0.8   &  0.39   &    20.4   &   5.1   &   1.3   &  0.60   &   u   &   3  \\ 
  51  &   197.86033   &  -1.32997   &    19.6   &    20.0   &   3.0   &   0.9   &  0.36   &    20.7   &   0.8   &   2.9   &  0.37   &   y   &   1  \\ 
  52  &   197.90840   &  -1.32801   &    19.4   &    19.7   &   3.4   &   3.7   &  0.59   &    21.0   &   2.8   &   1.5   &  0.60   &   u   &   1  \\ 
  53  &   197.87004   &  -1.35046   &    20.1   &    20.7   &   2.3   &   1.2   &  0.60   &    21.1   &   2.7   &   4.0   &  0.46   &   u   &   3  \\ 
  54  &   197.87458   &  -1.33752   &    19.0   &    20.6   &   3.8   &   0.5   &  0.58   &    19.2   &   2.5   &   4.0   &  0.30   &   y   &   2  \\ 
  55  &   197.86794   &  -1.33212   &    21.0   &    21.5   &   1.0   &   1.5   &  0.14   &    22.0   &   3.9   &   0.5   &  0.60   &   u   &   3  \\ 
  56  &   197.87754   &  -1.34546   &    17.8   &    18.1   &   2.1   &   3.1   &  0.07   &    19.2   &   5.1   &   0.7   &  0.30   &   y   &   2  \\ 
  57  &   197.87279   &  -1.32510   &    19.7   &    20.9   &   2.0   &   0.5   &  0.41   &    20.2   &   2.5   &   4.0   &  0.47   &   u   &   1  \\ 
  58  &   197.87144   &  -1.36533   &    18.1   &    18.6   &   2.4   &   3.7   &  0.38   &    19.3   &   5.1   &   1.6   &  0.15   &   y   &   1  \\ 
  59  &   197.87688   &  -1.31791   &    20.5   &    21.9   &   1.9   &   0.8   &  0.60   &    20.8   &   3.3   &   1.7   &  0.19   &   u   &   3  \\ 
  60  &   197.86766   &  -1.34539   &    17.6   &    17.7   &   7.2   &   2.3   &  0.17   &    19.7   &   0.6   &   2.0   &  0.22   &   y   &   2  \\ 
  61  &   197.86944   &  -1.35081   &    20.2   &    20.9   &   5.1   &   0.5   &  0.38   &    21.1   &   1.6   &   1.4   &  0.46   &   u   &   1  \\ 
  62  &   197.87163   &  -1.35573   &    20.8   &    22.0   &   0.7   &   2.2   &  0.06   &    21.3   &   2.3   &   0.6   &  0.11   &   u   &   1  \\ 
  63  &   197.87098   &  -1.35464   &    17.6   &    17.9   &   6.4   &   2.3   &  0.03   &    19.4   &   0.9   &   2.1   &  0.33   &   y   &   1  \\ 
  64  &   197.85024   &  -1.31872   &    18.7   &    19.0   &   5.4   &   6.0   &  0.42   &    20.1   &   5.1   &   2.0   &  0.59   &   y   &   1  \\ 
  65  &   197.87606   &  -1.34771   &    18.0   &    18.2   &   2.9   &   1.2   &  0.18   &    19.5   &   0.4   &   1.2   &  0.06   &   y   &   1  \\ 
  66  &   197.87121   &  -1.34472   &    20.5   &    21.7   &   1.0   &   2.0   &  0.22   &    20.9   &   2.3   &   1.2   &  0.12   &   u   &   1  \\ 
  67  &   197.87983   &  -1.35697   &    17.4   &    17.5   &   8.4   &   4.0   &  0.11   &    20.1   &   0.6   &   4.0   &  0.21   &   y   &   2  \\ 
  68  &   197.88386   &  -1.36057   &    17.9   &    19.6   &  11.6   &   0.8   &  0.60   &    18.1   &   2.8   &   4.0   &  0.25   &   y   &   2  \\ 
  69  &   197.85814   &  -1.33107   &    18.2   &    18.3   &   2.3   &   3.5   &  0.23   &    20.9   &   5.1   &   4.0   &  0.60   &   y   &   2  \\ 
  70  &   197.86637   &  -1.36018   &    18.7   &    18.9   &   3.0   &   1.1   &  0.60   &    20.4   &   0.3   &   1.5   &  0.17   &   y   &   3  \\ 
  71  &   197.87630   &  -1.34143   &    17.0   &    17.6   &   5.6   &   0.7   &  0.06   &    18.1   &   1.1   &   2.1   &  0.05   &   y   &   2  \\ 
  72  &   197.90779   &  -1.32237   &    17.1   &    17.3   &   4.5   &   5.1   &  0.05   &    18.9   &   3.4   &   4.0   &  0.27   &   y   &   2  \\ 
  73  &   197.89215   &  -1.35517   &    18.9   &    19.3   &   3.2   &   1.3   &  0.60   &    20.1   &   0.7   &   2.6   &  0.22   &   y   &   3  \\ 
  74  &   197.86295   &  -1.31333   &    19.9   &    21.9   &   3.2   &   0.5   &  0.40   &    20.1   &   2.6   &   3.3   &  0.28   &   y   &   1  \\ 
  75  &   197.86387   &  -1.31076   &    20.3   &    21.3   &   3.3   &   0.5   &  0.41   &    20.8   &   1.2   &   1.7   &  0.49   &   u   &   2  \\ 
  76  &   197.88723   &  -1.36132   &    19.6   &    19.7   &   2.5   &   2.9   &  0.43   &    24.4   &   1.7   &   0.5   &  0.60   &   u   &   1  \\ 
  77  &   197.86614   &  -1.35358   &    18.9   &    19.5   &   3.2   &   1.1   &  0.26   &    19.8   &   0.6   &   3.4   &  0.16   &   y   &   2  \\ 
  78  &   197.87653   &  -1.35320   &    20.3   &    21.3   &   1.1   &   1.8   &  0.27   &    20.9   &   2.2   &   0.8   &  0.30   &   u   &   2  \\ 
  79  &   197.86341   &  -1.34264   &    20.1   &    21.6   &   1.9   &   0.8   &  0.53   &    20.4   &   3.1   &   4.0   &  0.37   &   u   &   1  \\ 
  81  &   197.87489   &  -1.33220   &    18.5   &    18.6   &   2.0   &   2.0   &  0.06   &    20.9   &   0.4   &   1.9   &  0.06   &   y   &   1  \\ 
  82  &   197.84903   &  -1.34259   &    20.6   &    21.1   &   2.3   &   3.0   &  0.60   &    21.7   &   2.0   &   0.6   &  0.60   &   u   &   2  \\ 
  83  &   197.88260   &  -1.34962   &    18.5   &    18.7   &   4.1   &   6.0   &  0.15   &    20.7   &   5.0   &   1.0   &  0.42   &   y   &   1  \\ 
  84  &   197.85195   &  -1.35305   &    18.0   &    19.6   &   6.2   &   6.0   &  0.35   &    18.2   &   2.9   &   4.0   &  0.60   &   y   &   2  \\ 
  85  &   197.88627   &  -1.34956   &    19.8   &    20.4   &   3.1   &   1.4   &  0.37   &    20.7   &   1.0   &   3.5   &  0.47   &   u   &   1  \\ 
  86  &   197.89496   &  -1.34977   &    18.2   &    18.8   &   4.0   &   1.5   &  0.43   &    19.3   &   0.8   &   1.9   &  0.19   &   y   &   2  \\ 
  87  &   197.87208   &  -1.33173   &    19.9   &    20.2   &   3.8   &   2.2   &  0.51   &    21.6   &   0.7   &   2.1   &  0.37   &   u   &   1  \\ 
  88  &   197.88258   &  -1.32079   &    19.9   &    20.3   &   1.3   &   2.1   &  0.45   &    21.1   &   3.5   &   0.6   &  0.33   &   u   &   1  \\ 
  89  &   197.87876   &  -1.34360   &    21.0   &    22.5   &   1.0   &   3.3   &  0.47   &    21.3   &   1.3   &   0.7   &  0.16   &   u   &   2  \\ 
  90  &   197.87889   &  -1.34106   &    19.8   &    19.9   &   2.2   &   1.2   &  0.60   &    21.9   &   0.5   &   1.7   &  0.47   &   u   &   2  \\ 
  91  &   197.86058   &  -1.31772   &    19.4   &    19.8   &   2.1   &   5.5   &  0.34   &    20.7   &   2.3   &   0.7   &  0.52   &   y   &   2  \\ 
  92  &   197.88435   &  -1.33586   &    21.0   &    21.3   &   1.7   &   1.3   &  0.24   &    22.6   &   1.3   &   1.2   &  0.60   &   u   &   1  \\ 
  93  &   197.84260   &  -1.35661   &    19.0   &    19.5   &   1.0   &   6.0   &  0.40   &    20.0   &   2.2   &   0.5   &  0.31   &   u   &   2  \\ 
  94  &   197.89081   &  -1.32523   &    21.1   &    21.7   &   1.9   &   1.8   &  0.66   &    21.9   &   1.7   &   1.7   &  0.64   &   u   &   1  \\ 
  95  &   197.88344   &  -1.32341   &    18.1   &    20.5   &   1.0   &   6.0   &  0.26   &    18.2   &   2.7   &   3.6   &  0.23   &   y   &   1  \\ 
  96  &   197.89665   &  -1.35353   &    19.1   &    19.4   &   2.6   &   1.8   &  0.29   &    20.8   &   0.5   &   1.4   &  0.13   &   y   &   1  \\ 
  97  &   197.89070   &  -1.31860   &    21.1   &    21.5   &   2.9   &   1.6   &  0.58   &    22.3   &   1.0   &   1.4   &  0.60   &   u   &   1  \\ 
  98  &   197.86606   &  -1.33547   &    17.5   &    17.9   &   4.2   &   1.4   &  0.06   &    18.8   &   0.5   &   1.6   &  0.02   &   y   &   1  \\ 
  99  &   197.87315   &  -1.33999   &    21.1   &    21.1   &   1.5   &   1.8   &  0.14   &    -   &   -   &   -   &  -   &   u   &   1  \\ 
 100  &   197.89361   &  -1.37080   &    19.1   &    19.6   &  12.9   &   3.4   &  0.06   &    20.2   &   1.2   &   2.3   &  0.02   &   u   &   1  \\ 
 101  &   197.85506   &  -1.32141   &    20.6   &    21.2   &   1.3   &   1.3   &  0.01   &    21.4   &   2.7   &   4.0   &  0.06   &   u   &   1  \\ 
 102  &   197.87199   &  -1.36153   &    21.2   &    23.4   &   4.5   &   0.5   &  0.46   &    21.3   &   1.4   &   1.1   &  0.28   &   u   &   3  \\ 
 103  &   197.88728   &  -1.34736   &    20.5   &    20.8   &   1.7   &   1.9   &  0.37   &    22.0   &   3.3   &   4.0   &  0.28   &   u   &   1  \\ 
 104  &   197.88429   &  -1.36964   &    17.4   &    17.5   &  15.2   &   5.9   &  0.28   &    19.4   &   1.0   &   3.2   &  0.23   &   y   &   2  \\ 
 105  &   197.89068   &  -1.32743   &    20.0   &    20.7   &   3.9   &   6.0   &  0.60   &    20.7   &   1.6   &   1.0   &  0.60   &   n   &   2  \\ 
 106  &   197.88396   &  -1.32965   &    17.5   &    18.4   &   5.8   &   0.5   &  0.11   &    18.2   &   1.1   &   2.7   &  0.07   &   y   &   2  \\ 
 107  &   197.85240   &  -1.34852   &    20.0   &    20.2   &   2.0   &   2.5   &  0.18   &    21.9   &   0.9   &   2.1   &  0.29   &   u   &   1  \\ 
 108  &   197.86452   &  -1.32156   &    20.7   &    22.3   &   1.7   &   0.5   &  0.14   &    20.9   &   1.4   &   2.6   &  0.31   &   u   &   1  \\ 
 109  &   197.84447   &  -1.36221   &    20.9   &    21.0   &   1.4   &   1.8   &  0.44   &    22.9   &   2.8   &   0.5   &  0.21   &   u   &   1  \\ 
 110  &   197.87510   &  -1.36870   &    17.9   &    18.5   &   9.6   &   1.1   &  0.49   &    18.9   &   1.0   &   2.6   &  0.29   &   y   &   1  \\ 
 111  &   197.86220   &  -1.32689   &    17.8   &    18.5   &  14.8   &   0.5   &  0.54   &    18.6   &   1.7   &   4.0   &  0.21   &   y   &   2  \\ 
 112  &   197.90224   &  -1.32363   &    20.4   &    20.8   &   4.3   &   1.8   &  0.47   &    21.7   &   0.7   &   2.2   &  0.47   &   n   &   1  \\ 
 113  &   197.85781   &  -1.30941   &    19.8   &    20.6   &   2.4   &   1.1   &  0.04   &    20.5   &   0.7   &   3.4   &  0.00   &   u   &   1  \\ 
 114  &   197.86104   &  -1.33803   &    20.8   &    20.9   &   1.5   &   2.0   &  0.11   &    23.9   &   1.1   &   4.0   &  0.48   &   u   &   1  \\ 
 115  &   197.86573   &  -1.33767   &    20.3   &    20.9   &   1.9   &   0.6   &  0.26   &    21.1   &   0.6   &   2.3   &  0.37   &   u   &   1  \\ 
 116  &   197.88895   &  -1.32858   &    18.8   &    19.9   &   5.5   &   0.5   &  0.46   &    19.4   &   0.9   &   2.5   &  0.09   &   y   &   2  \\ 
 117  &   197.85933   &  -1.33234   &    18.1   &    18.3   &   1.6   &   2.9   &  0.41   &    19.9   &   5.1   &   2.0   &  0.15   &   y   &   1  \\ 
 118  &   197.88466   &  -1.31970   &    20.7   &    20.8   &   1.8   &   1.4   &  0.60   &    24.7   &   0.1   &   0.5   &  0.38   &   u   &   2  \\ 
 119  &   197.90482   &  -1.32604   &    20.9   &    21.1   &   1.8   &   1.9   &  0.23   &    22.7   &   0.9   &   0.5   &  0.29   &   u   &   1  \\ 
 120  &   197.87926   &  -1.35771   &    19.0   &    19.4   &   1.3   &   3.0   &  0.27   &    20.5   &   4.0   &   0.5   &  0.50   &   y   &   2  \\ 
 121  &   197.89499   &  -1.32348   &    18.6   &    18.8   &   1.7   &   4.4   &  0.12   &    20.6   &   5.1   &   0.5   &  0.28   &   y   &   2  \\ 
 122  &   197.86949   &  -1.34689   &    21.3   &    22.5   &   2.1   &   0.5   &  0.21   &    21.7   &   0.9   &   1.3   &  0.36   &   u   &   1  \\ 
 123  &   197.89743   &  -1.33805   &    20.9   &    21.5   &   2.1   &   3.0   &  0.60   &    21.9   &   1.5   &   1.7   &  0.59   &   u   &   1  \\ 
 124  &   197.87127   &  -1.35993   &    20.5   &    21.2   &   3.0   &   0.6   &  0.60   &    21.2   &   1.0   &   1.6   &  0.26   &   y   &   3  \\ 
 125  &   197.89716   &  -1.32137   &    21.2   &    22.1   &   1.0   &   1.8   &  0.43   &    21.9   &   2.5   &   1.6   &  0.60   &   u   &   1  \\ 
 127  &   197.89266   &  -1.32036   &    20.9   &    22.4   &   2.4   &   0.8   &  0.08   &    21.3   &   1.2   &   2.8   &  0.36   &   u   &   1  \\ 
 128  &   197.84925   &  -1.34023   &    19.8   &    19.8   &   2.2   &   4.9   &  0.03   &    24.5   &   0.5   &   0.5   &  0.49   &   u   &   1  \\ 
 129  &   197.89676   &  -1.35528   &    18.6   &    18.6   &   2.1   &   4.1   &  0.23   &    21.9   &   1.2   &   0.5   &  0.60   &   y   &   1  \\ 
 130  &   197.87844   &  -1.34186   &    18.0   &    18.2   &   7.3   &   6.0   &  0.15   &    19.8   &   1.3   &   3.1   &  0.50   &   y   &   2  \\ 
 131  &   197.86318   &  -1.31808   &    19.1   &    19.5   &   5.8   &   6.0   &  0.15   &    20.3   &   1.5   &   2.6   &  0.59   &   y   &   2  \\ 
 132  &   197.89754   &  -1.34519   &    18.3   &    18.7   &   3.5   &   3.1   &  0.09   &    19.6   &   0.6   &   4.0   &  0.04   &   y   &   1  \\ 
 133  &   197.85740   &  -1.34258   &    20.0   &    21.9   &   1.8   &   6.0   &  0.34   &    20.2   &   1.6   &   1.5   &  0.59   &   u   &   3  \\ 
 134  &   197.86424   &  -1.33401   &    18.1   &    18.9   &   6.0   &   1.2   &  0.05   &    18.9   &   1.0   &   4.0   &  0.24   &   y   &   1  \\ 
 135  &   197.87232   &  -1.31881   &    20.2   &    20.2   &   1.6   &   1.9   &  0.54   &    23.9   &   0.2   &   0.7   &  0.24   &   u   &   1  \\ 
 136  &   197.89539   &  -1.33456   &    19.1   &    19.9   &   3.4   &   0.6   &  0.38   &    19.8   &   0.5   &   3.0   &  0.29   &   y   &   1  \\ 
 137  &   197.87269   &  -1.33644   &    21.0   &    21.1   &   1.2   &   1.1   &  0.13   &    23.9   &   0.1   &   1.2   &  0.04   &   u   &   1  \\ 
 138  &   197.89822   &  -1.35255   &    19.0   &    20.0   &   4.2   &   0.5   &  0.13   &    19.6   &   0.8   &   2.8   &  0.26   &   y   &   2  \\ 
 139  &   197.87416   &  -1.31256   &    21.1   &    21.1   &   1.4   &   1.2   &  0.56   &    -   &   -   &   -   &  -   &   u   &   2  \\ 
 140  &   197.87964   &  -1.34792   &    18.5   &    18.9   &   2.2   &   1.0   &  0.08   &    19.7   &   0.3   &   1.6   &  0.03   &   y   &   1  \\ 
 142  &   197.86495   &  -1.34294   &    20.8   &    21.8   &   1.0   &   2.6   &  0.50   &    21.4   &   1.2   &   0.9   &  0.13   &   u   &   1  \\ 
 143  &   197.87410   &  -1.35017   &    18.9   &    19.2   &   1.0   &   3.8   &  0.13   &    20.6   &   3.9   &   0.5   &  0.24   &   y   &   1  \\ 
 144  &   197.87809   &  -1.36978   &    20.7   &    21.8   &   2.6   &   0.7   &  0.42   &    21.2   &   0.9   &   1.4   &  0.60   &   u   &   1  \\ 
 145  &   197.84153   &  -1.35485   &    21.2   &    21.4   &   1.0   &   1.7   &  0.33   &    22.9   &   2.2   &   0.5   &  0.16   &   u   &   1  \\ 
 146  &   197.86722   &  -1.32543   &    18.4   &    18.8   &   1.0   &   4.1   &  0.25   &    19.7   &   5.1   &   0.8   &  0.60   &   y   &   2  \\ 
 148  &   197.87969   &  -1.35771   &    18.1   &    18.2   &   1.6   &   3.3   &  0.24   &    21.1   &   4.6   &   0.5   &  0.60   &   y   &   2  \\ 
 149  &   197.86556   &  -1.35226   &    20.8   &    22.1   &   1.2   &   0.5   &  0.01   &    21.3   &   0.6   &   1.7   &  0.29   &   y   &   3  \\ 
 150  &   197.88038   &  -1.36321   &    21.0   &    21.1   &   1.4   &   1.7   &  0.48   &    24.8   &   0.3   &   3.6   &  0.60   &   u   &   1  \\ 
 151  &   197.89948   &  -1.33262   &    19.4   &    19.5   &   2.3   &   3.6   &  0.36   &    21.8   &   0.7   &   0.5   &  0.49   &   y   &   1  \\ 
 152  &   197.86628   &  -1.36595   &    19.6   &    19.7   &   1.6   &   1.7   &  0.60   &    21.9   &   0.3   &   1.9   &  0.60   &   u   &   1  \\ 
 153  &   197.84824   &  -1.35306   &    21.1   &    24.5   &   3.8   &   0.5   &  0.37   &    21.2   &   1.3   &   1.5   &  0.60   &   u   &   2  \\ 
 154  &   197.87287   &  -1.36927   &    21.3   &    21.4   &   1.0   &   1.5   &  0.43   &    23.8   &   2.2   &   0.5   &  0.27   &   u   &   2  \\ 
 155  &   197.85585   &  -1.33810   &    18.6   &    19.3   &  10.5   &   0.8   &  0.34   &    19.4   &   1.0   &   4.0   &  0.16   &   y   &   2  \\ 
 156  &   197.85475   &  -1.32528   &    18.6   &    18.8   &   5.3   &   3.5   &  0.09   &    20.2   &   0.4   &   4.0   &  0.41   &   y   &   1  \\ 
 157  &   197.88115   &  -1.35167   &    20.7   &    20.8   &   1.2   &   2.0   &  0.39   &    24.0   &   0.6   &   4.0   &  0.60   &   y   &   1  \\ 
 158  &   197.87148   &  -1.31401   &    20.5   &    20.9   &   5.7   &   6.0   &  0.23   &    22.0   &   1.3   &   2.6   &  0.30   &   u   &   1  \\ 
 159  &   197.87676   &  -1.34301   &    19.1   &    20.3   &   3.1   &   0.8   &  0.47   &    19.5   &   0.8   &   2.0   &  0.40   &   y   &   2  \\ 
 160  &   197.86885   &  -1.34174   &    20.4   &    21.3   &   1.0   &   2.1   &  0.42   &    21.1   &   2.6   &   2.8   &  0.60   &   u   &   2  \\ 
 161  &   197.87671   &  -1.34594   &    18.1   &    21.8   &   1.0   &   2.4   &  0.37   &    18.2   &   1.7   &   4.0   &  0.14   &   y   &   1  \\ 
 162  &   197.84011   &  -1.36156   &    19.6   &    20.4   &   2.3   &   1.3   &  0.13   &    20.3   &   0.5   &   4.0   &  0.13   &   y   &   1  \\ 
 163  &   197.87094   &  -1.34828   &    18.7   &    18.8   &   1.1   &   2.8   &  0.24   &    21.6   &   1.5   &   0.5   &  0.47   &   y   &   2  \\ 
 164  &   197.87850   &  -1.34863   &    20.5   &    20.8   &   1.3   &   2.1   &  0.31   &    22.0   &   1.1   &   0.7   &  0.60   &   y   &   2  \\ 
 165  &   197.86525   &  -1.36081   &    20.1   &    21.3   &   2.8   &   1.2   &  0.09   &    20.5   &   1.0   &   1.8   &  0.60   &   u   &   2  \\ 
 166  &   197.87250   &  -1.34941   &    20.6   &    21.9   &   1.2   &   0.8   &  0.37   &    21.1   &   0.9   &   2.1   &  0.33   &   u   &   2  \\ 
 167  &   197.87493   &  -1.31782   &    20.3   &    20.9   &   1.3   &   2.1   &  0.30   &    21.2   &   0.8   &   1.0   &  0.22   &   u   &   2  \\ 
 168  &   197.85088   &  -1.35207   &    18.8   &    19.5   &   2.8   &   1.0   &  0.12   &    19.5   &   0.4   &   2.9   &  0.14   &   y   &   2  \\ 
 169  &   197.89110   &  -1.33224   &    20.1   &    20.6   &   1.0   &   4.1   &  0.21   &    21.2   &   1.9   &   1.7   &  0.59   &   u   &   2  \\ 
 170  &   197.87456   &  -1.32530   &    20.9   &    22.0   &   1.8   &   1.1   &  0.07   &    21.4   &   0.8   &   1.9   &  0.56   &   u   &   1  \\ 
 171  &   197.87530   &  -1.36194   &    18.1   &    18.9   &  13.5   &   5.6   &  0.22   &    18.8   &   1.1   &   3.6   &  0.60   &   y   &   1  \\ 
 172  &   197.87222   &  -1.32127   &    18.5   &    20.9   &   7.9   &   0.5   &  0.60   &    18.6   &   1.3   &   3.0   &  0.60   &   y   &   3  \\ 
 173  &   197.87513   &  -1.34275   &    20.5   &    21.1   &   1.7   &   0.6   &  0.57   &    21.4   &   0.4   &   2.0   &  0.52   &   u   &   2  \\ 
 174  &   197.87526   &  -1.35504   &    19.0   &    20.1   &   1.0   &   6.0   &  0.22   &    19.5   &   1.1   &   1.7   &  0.37   &   y   &   2  \\ 
 175  &   197.87668   &  -1.35269   &    20.2   &    20.5   &   4.2   &   6.0   &  0.47   &    21.6   &   0.7   &   4.0   &  0.52   &   u   &   3  \\ 
 176  &   197.86865   &  -1.34030   &    18.1   &    18.2   &   3.7   &   6.0   &  0.08   &    20.6   &   0.5   &   0.8   &  0.17   &   y   &   2  \\ 
 177  &   197.87036   &  -1.33387   &    21.1   &    21.6   &   1.2   &   1.3   &  0.08   &    22.2   &   0.6   &   2.1   &  0.50   &   y   &   1  \\ 
 178  &   197.88933   &  -1.32345   &    20.3   &    21.2   &   1.5   &   0.8   &  0.17   &    21.0   &   0.3   &   4.0   &  0.16   &   u   &   2  \\ 
 179  &   197.87228   &  -1.35152   &    19.5   &    20.7   &   1.2   &   0.8   &  0.28   &    19.9   &   0.7   &   4.0   &  0.21   &   y   &   1  \\ 
 180  &   197.86765   &  -1.36779   &    20.9   &    21.2   &   1.0   &   1.4   &  0.27   &    22.7   &   0.5   &   1.2   &  0.60   &   y   &   1  \\ 
 181  &   197.87367   &  -1.36065   &    21.2   &    22.2   &   1.8   &   1.1   &  0.22   &    21.7   &   0.6   &   1.5   &  0.50   &   u   &   1  \\ 
 182  &   197.87992   &  -1.34560   &    19.4   &    20.5   &   2.4   &   0.7   &  0.07   &    19.8   &   0.4   &   2.0   &  0.10   &   y   &   2  \\ 
 183  &   197.87516   &  -1.33814   &    19.2   &    19.7   &   6.9   &   4.2   &  0.60   &    20.1   &   0.4   &   3.1   &  0.26   &   y   &   1  \\ 
 184  &   197.88871   &  -1.36590   &    20.8   &    21.1   &   1.2   &   4.9   &  0.07   &    22.4   &   0.5   &   1.3   &  0.10   &   u   &   1  \\ 
 185  &   197.89166   &  -1.37136   &    20.3   &    20.6   &   1.0   &   2.0   &  0.16   &    21.9   &   0.2   &   3.4   &  0.22   &   u   &   1  \\ 
 186  &   197.87828   &  -1.31459   &    20.1   &    20.9   &   1.3   &   1.6   &  0.33   &    20.9   &   0.4   &   3.4   &  0.60   &   y   &   1  \\ 
 187  &   197.87125   &  -1.32970   &    19.4   &    19.8   &   1.4   &   1.7   &  0.60   &    20.6   &   0.2   &   3.5   &  0.11   &   y   &   2  \\ 
 188  &   197.87534   &  -1.36657   &    20.5   &    20.8   &   1.1   &   6.0   &  0.05   &    22.3   &   0.5   &   1.1   &  0.51   &   u   &   2  \\ 
 189  &   197.87044   &  -1.33282   &    21.4   &    22.3   &   1.0   &   0.7   &  0.11   &    22.0   &   0.3   &   3.5   &  0.08   &   u   &   1  \\ 
 190  &   197.87226   &  -1.34055   &    20.2   &    21.4   &   1.0   &   6.0   &  0.50   &    20.6   &   0.6   &   1.2   &  0.14   &   u   &   2  \\ 
 191  &   197.88846   &  -1.32137   &    20.3   &    21.0   &   1.0   &   6.0   &  0.05   &    21.1   &   0.5   &   1.5   &  0.11   &   y   &   1  \\ 
 192  &   197.87580   &  -1.34155   &    20.1   &    20.7   &   3.5   &   0.6   &  0.60   &    20.9   &   0.4   &   2.4   &  0.07   &   y   &   2  \\ 
 193  &   197.88103   &  -1.31603   &    21.0   &    21.8   &   1.0   &   6.0   &  0.42   &    21.7   &   0.5   &   1.3   &  0.15   &   u   &   1  \\ 
 194  &   197.87643   &  -1.34060   &    20.8   &    21.4   &   1.1   &   6.0   &  0.54   &    21.6   &   0.3   &   2.0   &  0.24   &   y   &   1  \\
 cD   &   197.87297   &  -1.34110   &    15.7   &    15.7   &   8.3   &   7.0   &  -     &     -     &    -    &    -    &    -    &   y   &   1   \\
 
%cD   &   197.87297   &  -1.34110   &    15.7   &    15.7   &   8.3   &   7.0   $   -     &     -     &    -    &    -    &    -    &   y   &   1\epsilon^{\rm c2}$ &  cluster  &  flag
\enddata
\tablecomments{Best fit parameters for the 2-D double component \ser models for the first 179 galaxies. The ${R_e}$'s are major-axis values. For galaxies \#99 and \#139 the second component had null contribution. The cD parameters are the 1-D single \csersic\ fit, with ${R_b}=$3.8~kpc, $\gamma=0.04$ and $\alpha$=2.6; ${R_e}$ and ${R_b}$ are circularized values. 
The cluster member column indicates if the galaxy is a confirmed member (y), known non-member (n), or unknown (u). \\
Flag 1 (good) and 2 (acceptable) are reliable fits, while 3 indicates the fit is unreliable. }
\end{deluxetable}

\clearpage

\startlongtable
\begin{deluxetable*}{*{5}{c}}
\tablecaption{Globular cluster system \ser parameters
\label{globular.tab}}  
\tablehead{ ${\rm ID}$ & $N_{\rm GC}$  & $R_e^{\rm GC}$ [kpc] & \ser index & $S_N$ }
\startdata
   5  &    4142 $\pm$    284  &   19.0 &  4.0 &    5.4 $\pm$    1.1   \\ 
  29  &    4858 $\pm$    307  &   29.5 &  4.0 &    4.1 $\pm$    0.8   \\ 
  33  &    4858 $\pm$    307  &   33.1 &  4.0 &    4.0 $\pm$    0.8   \\ 
  51  &      50 $\pm$     31  &    8.8 &  4.0 &    0.8 $\pm$    0.6   \\ 
  56  &     706 $\pm$    117  &    3.4 &  1.6 &    2.2 $\pm$    0.7   \\ 
  58  &    1337 $\pm$    162  &   13.0 &  0.7 &    5.9 $\pm$    1.5   \\ 
  60  &    1221 $\pm$    154  &    8.6 &  4.0 &    3.2 $\pm$    0.8   \\ 
  63  &    4858 $\pm$    307  &   21.1 &  2.2 &   13.3 $\pm$    2.7   \\ 
  65  &    2032 $\pm$    199  &    9.1 &  1.5 &    7.7 $\pm$    1.8   \\ 
  67  &    3408 $\pm$    257  &   30.9 &  4.0 &    7.8 $\pm$    1.7   \\ 
  68  &    1218 $\pm$    155  &   18.5 &  4.0 &    4.2 $\pm$    1.1   \\ 
  69  &     881 $\pm$    132  &   10.3 &  4.0 &    4.2 $\pm$    1.2   \\ 
  72  &     415 $\pm$     91  &   17.1 &  4.0 &    0.7 $\pm$    0.3   \\ 
  86  &     600 $\pm$    109  &   11.0 &  4.0 &    2.9 $\pm$    0.9   \\ 
  96  &     405 $\pm$     90  &    5.7 &  0.9 &    4.5 $\pm$    1.6   \\ 
  98  &    2529 $\pm$    221  &   12.4 &  4.0 &    6.3 $\pm$    1.4   \\ 
 106  &    2086 $\pm$    201  &   12.6 &  4.0 &    5.3 $\pm$    1.3   \\ 
 111  &    1924 $\pm$    194  &   33.6 &  2.3 &    6.3 $\pm$    1.5   \\ 
 116  &     360 $\pm$     84  &    8.6 &  4.0 &    3.0 $\pm$    1.1   \\ 
 117  &     301 $\pm$     78  &    1.1 &  0.5 &    1.3 $\pm$    0.5   \\ 
 120  &      50 $\pm$     31  &    8.0 &  4.0 &    0.5 $\pm$    0.4   \\ 
 121  &     142 $\pm$     55  &    8.8 &  4.0 &    0.9 $\pm$    0.5   \\ 
 129  &     449 $\pm$     95  &    8.2 &  4.0 &    3.0 $\pm$    1.0   \\ 
 132  &     369 $\pm$     87  &   10.5 &  1.5 &    1.9 $\pm$    0.7   \\ 
 134  &    2362 $\pm$    214  &   13.9 &  4.0 &   10.4 $\pm$    2.4   \\ 
 140  &     538 $\pm$    103  &    6.0 &  0.9 &    3.3 $\pm$    1.1   \\ 
 148  &      50 $\pm$     31  &    6.7 &  1.8 &    0.2 $\pm$    0.2   \\ 
 151  &     171 $\pm$     60  &    8.4 &  4.0 &    2.4 $\pm$    1.2   \\ 
 155  &     150 $\pm$     56  &    3.0 &  4.0 &    1.0 $\pm$    0.5   \\ 
 156  &    2131 $\pm$    204  &   16.6 &  3.4 &   14.0 $\pm$    3.3   \\ 
 161  &    1461 $\pm$    168  &    6.7 &  4.0 &    6.4 $\pm$    1.6   \\ 
 171  &    1681 $\pm$    181  &   28.6 &  1.3 &    7.4 $\pm$    1.8   \\ 
  cD  &   39574 $\pm$   11872 &   39.7 &  1.0 &   18.0 $\pm$    9.0   \\   
 ICL  &   53679 $\pm$   16104 &  172.0 &  1.0 &   16.9 $\pm$    6.0   \\
 
\enddata
\tablecomments{Best 2-D single \ser models for the confirmed A1689 members and classified as reliable fits. $N_{\rm GC}$ is the total number after correcting for incompleteness, background contamination, uncovered areas, and GCLF. The values for the cD and ICL are from 1-D \csersic\ and \ser, respectively.  }
\end{deluxetable*}


\newpage

\begin{thebibliography}{}
\bibitem[Abell(1965)]{1965ARA&A...3....1A} Abell, G.~O.\ 1965, \araa, 3, 1 
\bibitem[Alamo-Mart{\'{\i}}nez et al.(2013)]{2013ApJ...775...20A} Alamo-Mart{\'{\i}}nez, K.~A., Blakeslee, J.~P., Jee, M.~J., et al.\ 2013, \apj, 775, 20~(AM13)
\bibitem[Anderson \& Bedin(2010)]{2010PASP..122.1035A} Anderson, J., \& Bedin, L.~R.\ 2010, \pasp, 122, 1035 

% Galaxy bimodality versus stellar mass and environment
\bibitem[Baldry et al.(2006)]{2006MNRAS.373..469B} Baldry, I.~K., Balogh, M.~L., Bower,  R.~G., et al.\ 2006, \mnras, 373, 469  
\bibitem[Balogh et al.(2016)]{2016MNRAS.456.4364B} Balogh, M.~L., McGee, S.~L., Mok, A., et al.\ 2016, \mnras, 456, 4364 
\bibitem[Beers et al.(1990)]{1990AJ....100...32B} Beers, T.~C., Flynn, K., \& Gebhardt, K.\ 1990, \aj, 100, 32 
\bibitem[Begelman et al.(1980)]{1980Natur.287..307B} Begelman, M.~C., Blandford, R.~D., \& Rees, M.~J.\ 1980, \nat, 287, 307
\bibitem[Bender(1988)]{1988A&A...193L...7B} Bender, R.\ 1988, \aap, 193, L7 
\bibitem[Bender et al.(1989)]{1989A&A...217...35B} Bender, R., Surma, P., Doebereiner, S., Moellenhoff, C., \& Madejsky, R.\ 1989, \aap, 217, 35
\bibitem[Bender et al.(2015)]{2015ApJ...807...56B} Bender, R., Kormendy, J., Cornell, M.~E., \& Fisher, D.~B.\ 2015, \apj, 807, 56 
\bibitem[Bertin \& Arnouts(1996)]{1996A&AS..117..393B} Bertin, E., \& Arnouts, S.\ 1996, \aaps, 117, 393 
\bibitem[Blakeslee(1999)]{1999AJ....118.1506B} Blakeslee, J.~P.\ 1999, \aj, 118, 1506 
%\bibitem[Blakeslee(2005)]{2005HiA....13..171B} Blakeslee, J.~P.\ 2005, Highlights of
%Astronomy, 13, 171    
\bibitem[Blakeslee et al.(2015)]{2015AAS...22525209B} Blakeslee, J.~P., Alamo-Martinez,
  K., Toloba, E., Barro, G., \& Peng, E.~W.\ 2015, American Astronomical Society Meeting
  Abstracts, 225, 252.09  
\bibitem[Blakeslee et al.(2003)]{2003ASPC..295..257B} Blakeslee, J.~P., Anderson, K.~R., Meurer, G.~R.\ et al.\ 2003, {\it Astronomical Data Analysis Software and Systems XII} (ASP Conf. Ser. 295), ed. H. E. Payne, R. I. Jedrzejewski, \& R. N. Hook (San Francisco, CA: ASP), 257 
\bibitem[Blakeslee et al.(2006)]{2006ApJ...644...30B} Blakeslee, J.~P., Holden, B.~P., Franx, M., et al.\ 2006, \apj, 644, 30 
\bibitem[Blakeslee et al.(1997)]{1997AJ....114..482B} Blakeslee, J.~P., 
Tonry, J.~L., \& Metzger, M.~R.\ 1997, \aj, 114, 482 
\bibitem[Bluck et al.(2016)]{2016MNRAS.462.2559B} Bluck, A.~F.~L., Mendel, J.~T., Ellison, S.~L., et al.\ 2016, \mnras, 462, 2559 
% \bibitem[Bonfini et al.(2015)]{2015ApJ...807..136B} Bonfini, P., Dullo, B.~T., \& Graham, A.~W.\ 2015, \apj, 807, 136
\bibitem[Bonfini \& Graham(2016)]{2016ApJ...829...81B} Bonfini, P., \& Graham, A.~W.\ 2016, \apj, 829, 81
\bibitem[Boschin et al.(2006)]{2006A&A...449..461B} Boschin, W., Girardi, M., Spolaor, M., \& Barrena, R.\ 2006, \aap, 449, 461

\bibitem[Boylan-Kolchin et al.(2006)]{2006MNRAS.369.1081B} Boylan-Kolchin, M., Ma, C.-P., \& Quataert, E.\ 2006, \mnras, 369, 1081
\bibitem[Brodie \& Strader(2006)]{2006ARA&A..44..193B} Brodie, J.~P., \& Strader, J.\ 2006, \araa, 44, 193
\bibitem[Burkert \& Tremaine(2010)]{2010ApJ...720..516B} Burkert, A., \& Tremaine, S.\ 2010, \apj, 720, 516

\bibitem[Cappellari et al.(2011)]{2011MNRAS.416.1680C} Cappellari, M., Emsellem, E., Krajnovi{\'c}, D., et al.\ 2011, \mnras, 416, 1680 
% VLA Imaging of Virgo Spirals in Atomic Gas (VIVA). I. The Atlas and the H I Properties
\bibitem[Cho et al.(2016)]{2016ApJ...822...95C} Cho, H., Blakeslee, J.~P., Chies-Santos, A.~L., et al.\ 2016, \apj, 822, 95 
\bibitem[Chung et al.(2009)]{2009AJ....138.1741C} Chung, A., van Gorkom, J.~H., Kenney, J.~D., Crowl, H., \& Vollmer, B.\ 2009, \aj, 138, 1741  
\bibitem[C{\^o}t{\'e} et al.(2004)]{2004ApJS..153..223C} C{\^o}t{\'e}, P., Blakeslee, J.~P., Ferrarese, L., et al.\ 2004, \apjs, 153, 223
\bibitem[Contini et al.(2014)]{2014MNRAS.437.3787C} Contini, E., De Lucia, G., Villalobos, {\'A}., \& Borgani, S.\ 2014, \mnras, 437, 3787
\bibitem[Cooper et al.(2015)]{2015MNRAS.451.2703C} Cooper, A.~P., Gao, L., Guo, Q., et al.\ 2015, \mnras, 451, 2703 

\bibitem[D'Eugenio et al.(2013)]{2013MNRAS.429.1258D} D'Eugenio, F., Houghton, R.~C.~W., Davies, R.~L.,  \& Dalla Bont{\`a}, E.\ 2013, \mnras, 429, 1258 
\bibitem[Donzelli et al.(2011)]{2011ApJS..195...15D} Donzelli, C.~J., Muriel, H., \& Madrid, J.~P.\ 2011, \apjs, 195, 15
\bibitem[Dressler(1980)]{1980ApJ...236..351D} Dressler, A.\ 1980, \apj, 236, 351
\bibitem[Durrell et al.(2014)]{2014ApJ...794..103D} Durrell, P.~R., C{\^o}t{\'e}, P., Peng, E.~W., et al.\ 2014, \apj, 794, 103 

\bibitem[Ebisuzaki et al.(1991)]{1991Natur.354..212E} Ebisuzaki, T., Makino, J., \& Okumura, S.~K.\ 1991, \nat, 354, 212 
\bibitem[Einasto et al.(1974)]{1974Natur.252..111E} Einasto, J., Saar, E., Kaasik, A., \& Chernin, A.~D.\ 1974, \nat, 252, 111 

\bibitem[Emsellem et al.(2007)]{2007MNRAS.379..401E} Emsellem, E., Cappellari, M., Krajnovi{\'c}, D., et al.\ 2007, \mnras, 379, 401 

\bibitem[Emsellem et al.(2011)]{2011MNRAS.414..888E} Emsellem, E., Cappellari, M., Krajnovi{\'c}, D., et al.\ 2011, \mnras, 414, 888

\bibitem[Ferrarese et al.(2006)]{2006ApJS..164..334F} Ferrarese, L.,
  C{\^o}t{\'e}, P., Jord{\'a}n, A., et al.\ 2006, \apjs, 164, 334 

\bibitem[Ferrarese \& Ford(2005)]{2005SSRv..116..523F} Ferrarese, L., \& Ford, H.\ 2005, \ssr, 116, 523 

\bibitem[Fossati et al.(2017)]{2017ApJ...835..153F} Fossati, M., Wilman, D.~J., Mendel, J.~T., et al.\ 2017, \apj, 835, 153 

\bibitem[Gallagher \& Ostriker(1972)]{1972AJ.....77..288G} Gallagher, J.~S., III, \& Ostriker, J.~P.\ 1972, \aj, 77, 288

\bibitem[Georgiev et al.(2010)]{2010MNRAS.406.1967G} Georgiev, I.~Y., Puzia, T.~H., Goudfrooij, P., \& Hilker, M.\ 2010, \mnras, 406, 1967 
\bibitem[Gonzalez et al.(2005)]{2005ApJ...618..195G} Gonzalez, A.~H., Zabludoff, A.~I., \& Zaritsky, D.\ 2005, \apj, 618, 195
\bibitem[Graham et al.(2003)]{2003AJ....125.2951G} Graham, A.~W., Erwin, P., Trujillo, I., \& Asensio Ramos, A.\ 2003, \aj, 125, 2951
\bibitem[Graham \& Driver(2005)]{2005PASA...22..118G} Graham, A.~W., \& Driver, S.~P.\ 2005, PASA, 22, 118 

\bibitem[Halkola et al.(2006)]{2006MNRAS.372.1425H} Halkola, A., Seitz, S., \& Pannella,
  M.\ 2006, \mnras, 372, 1425 (H06)

\bibitem[Harris(1991)]{1991ARA&A..29..543H} Harris, W.~E.\ 1991, \araa, 29, 543
\bibitem[Harris(2001)]{2001segc.book..223H} Harris, W.~E.\ 2001, in Star Clusters,
  Saas-Fee Advanced Courses, Vol.~28, ed.~L.~Labhardt \& B.~Binggeli (Berlin: Springer),~223
%ISBN 978-3-540-67646-1.
\bibitem[Harris et al.(2017)]{2017ApJ...836...67H} Harris, W.~E., Blakeslee, J.~P., \& Harris, G.~L.~H.\ 2017, \apj, 836, 67 

\bibitem[Harris et al.(2014)]{2014MNRAS.438.2117H} Harris, G.~L.~H., Poole, G.~B., \& Harris, W.~E.\ 2014, \mnras, 438, 2117 

\bibitem[Harris et al.(2013)]{2013ApJ...772...82H} Harris, W.~E., Harris, G.~L.~H., \&
  Alessi, M.\ 2013, \apj, 772, 82
%  
% \bibitem[Harris et al.(1995)]{1995ApJ...441..120H} Harris, W.~E., Pritchet, 
% C.~J., \& McClure, R.~D.\ 1995, \apj, 441, 120 
% \bibitem[Harris \& Racine(1979)]{1979ARA&A..17..241H} Harris, W.~E., \& Racine,
% R.\ 1979, \araa, 17, 241
%  
\bibitem[Harris \& van den Bergh(1981)]{1981AJ.....86.1627H} Harris, W.~E., \& van den Bergh, S.\ 1981, \aj, 86, 1627 
\bibitem[Hudson et al.(2015)]{2015MNRAS.447..298H} Hudson, M.~J., Gillis, B.~R., Coupon, J., et al.\ 2015, \mnras, 447, 298 
\bibitem[Hudson et al.(2014)]{2014ApJ...787L...5H} Hudson, M.~J., Harris, G.~L., \& Harris, W.~E.\ 2014, \apjl, 787, L5 

% BUDHIES - III: the fate of H I and the quenching of galaxies in evolving environments  
\bibitem[Jaff{\'e} et al.(2016)]{2016MNRAS.461.1202J} Jaff{\'e}, Y.~L., Verheijen, M.~A.~W., Haines, C.~P., et al.\ 2016, \mnras, 461, 1202  
\bibitem[Jord{\'a}n et al.(2003)]{2003AJ....125.1642J} Jord{\'a}n, A., West, M.~J., C{\^o}t{\'e}, P., \& Marzke, R.~O.\ 2003, \aj, 125, 1642

\bibitem[Kauffmann et al.(2004)]{2004MNRAS.353..713K} Kauffmann, G., White, S.~D.~M., Heckman, T.~M., et al.\ 2004, \mnras, 353, 713
\bibitem[Knobel et al.(2015)]{2015ApJ...800...24K} Knobel, C., Lilly, S.~J., Woo, J., \& Kova{\v c}, K.\ 2015, \apj, 800, 24
\bibitem[Ko et al.(2017)]{2017ApJ...835..212K} Ko, Y., Hwang, H.~S., Lee, M.~G., et al.\ 2017, \apj, 835, 212 
  
\bibitem[Komatsu et al.(2011)]{2011ApJS..192...18K} Komatsu, E., Smith, K.~M., Dunkley, J., et al.\ 2011, \apjs, 192, 18
\bibitem[Kormendy et al.(2009)]{2009ApJS..182..216K} Kormendy, J., Fisher, D.~B., Cornell, M.~E., \& Bender, R.\ 2009, \apjs, 182, 216 
\bibitem[Kormendy \& Ho(2013)]{2013ARA&A..51..511K} Kormendy, J., \& Ho, L.~C.\ 2013, \araa, 51, 511
\bibitem[Kravtsov \& Gnedin(2005)]{2005ApJ...623..650K} Kravtsov, A.~V., \& Gnedin, O.~Y.\ 2005, \apj, 623, 650
\bibitem[Kulkarni \& Loeb(2012)]{2012MNRAS.422.1306K} Kulkarni, G., \& Loeb, A.\ 2012, \mnras, 422, 1306 

\bibitem[Lagos et al.(2011)]{2011MNRAS.416.1566L} Lagos, C.~D.~P., Lacey, C.~G., Baugh, C.~M., Bower, R.~G., \& Benson, A.~J.\ 2011, \mnras, 416, 1566 
\bibitem[Lee \& Jang(2016)]{2016ApJ...831..108L} Lee, M.~G., \& Jang, I.~S.\ 2016, \apj, 831, 108 (LJ16)
\bibitem[Liu et al.(2016)]{2016ApJ...818..179L} Liu, Y., Peng, E.~W., Blakeslee, J., et al.\ 2016, \apj, 818, 179
\bibitem[L{\'o}pez-Cruz et al.(2014)]{2014ApJ...795L..31L} L{\'o}pez-Cruz, O., A{\~n}orve, C., Birkinshaw, M., et al.\ 2014, \apjl, 795, L31
\bibitem[Lotz et al.(2017)]{2017ApJ...837...97L} Lotz, J.~M., Koekemoer, A., Coe, D., et
  al.\ 2017, \apj, 837, 97  

\bibitem[Mackey et al.(2016)]{2016MNRAS.460L.114M} Mackey, A.~D., Beasley, M.~A., \& Leaman, R.\ 2016, \mnras, 460, L114 
\bibitem[Madrid \& Donzelli(2016)]{2016ApJ...819...50M} Madrid, J.~P., \& Donzelli, C.~J.\ 2016, \apj, 819, 50 
% Galaxy halo masses and satellite fractions from galaxy-galaxy lensing in the Sloan
% Digital Sky Survey: stellar mass, luminosity, morphology and environment dependencies 
\bibitem[Mandelbaum et al.(2006)]{2006MNRAS.368..715M} Mandelbaum, R., Seljak, U., Kauffmann, G., Hirata, C.~M., \& Brinkmann, J.\ 2006, \mnras, 368, 715

\bibitem[Matthews et al.(1964)]{1964ApJ...140...35M} Matthews, T.~A., Morgan, W.~W., \& Schmidt, M.\ 1964, \apj, 140, 35

\bibitem[McConnell \& Ma(2013)]{2013ApJ...764..184M} McConnell, N.~J., \& Ma, C.-P.\ 2013, \apj, 764, 184 

\bibitem[McConnell et al.(2011)]{2011Natur.480..215M} McConnell, N.~J., Ma, C.-P., Gebhardt, K., et al.\ 2011, \nat, 480, 215
\bibitem[McNamara et al.(2009)]{2009ApJ...698..594M} McNamara, B.~R., Kazemzadeh, F., Rafferty, D.~A., et al.\ 2009, \apj, 698, 594 
\bibitem[Merritt \& Milosavljevi{\'c}(2005)]{2005LRR.....8....8M} Merritt, D., \& Milosavljevi{\'c}, M.\ 2005, Living Reviews in Relativity, 8
\bibitem[Merritt(2006)]{2006ApJ...648..976M} Merritt, D.\ 2006, \apj, 648, 976
\bibitem[Mihos et al.(2005)]{2005ApJ...631L..41M} Mihos, J.~C., Harding, P., Feldmeier, J., \& Morrison, H.\ 2005, \apjl, 631, L41 
\bibitem[Mihos et al.(2017)]{2017ApJ...834...16M} Mihos, J.~C., Harding, P., Feldmeier, J.~J., et al.\ 2017, \apj, 834, 16 

\bibitem[Mistani et al.(2016)]{2016MNRAS.455.2323M} Mistani, P.~A., Sales, L.~V., Pillepich, A., et al.\ 2016, \mnras, 455, 2323
\bibitem[Mok et al.(2013)]{2013MNRAS.431.1090M} Mok, A., Balogh, M.~L., McGee, S.~L., et al.\ 2013, \mnras, 431, 1090 

\bibitem[Moster et al.(2010)]{2010ApJ...710..903M} Moster, B.~P., Somerville, R.~S., Maulbetsch, C., et al.\ 2010, \apj, 710, 903 

% The Gemini Cluster Astrophysics Spectroscopic Survey (GCLASS): The Role of Environment and Self-regulation in Galaxy Evolution at z ~ 1
\bibitem[Muzzin et al.(2012)]{2012ApJ...746..188M} Muzzin, A., Wilson, G., Yee, H.~K.~C., et al.\ 2012, \apj, 746, 188  

\bibitem[Naab et al.(2009)]{2009ApJ...699L.178N} Naab, T., Johansson, P.~H., \& Ostriker, J.~P.\ 2009, \apjl, 699, L178

\bibitem[Peng et al.(2009)]{2009ApJ...703...42P} Peng, E.~W., Jord{\'a}n, A., Blakeslee, J.~P., et al.\ 2009, \apj, 703, 42 
\bibitem[Peng et al.(2008)]{2008ApJ...681..197P} Peng, E.~W., Jord{\'a}n, A., C{\^o}t{\'e}, P., et al.\ 2008, \apj, 681, 197 
\bibitem[Peng et al.(2011)]{2011ApJ...730...23P} Peng, E.~W., Ferguson, H.~C.,  Goudfrooij, P., et al.\ 2011, \apj, 730, 23

% Mass and Environment as Drivers of Galaxy Evolution in SDSS and zCOSMOS and the Origin of the Schechter Function
\bibitem[Peng et al.(2010)]{2010ApJ...721..193P} Peng, Y.-j., Lilly, S.~J., Kova{\v c}, K., et al.\ 2010, \apj, 721, 193

  % Mass and Environment as Drivers of Galaxy Evolution. II. The Quenching of Satellite Galaxies as the Origin of Environmental Effects   
\bibitem[Peng et al.(2012)]{2012ApJ...757....4P} Peng, Y.-j., Lilly, S.~J., Renzini, A., \& Carollo, M.\ 2012, \apj, 757, 4  
  
\bibitem[Postman et al.(2005)]{2005ApJ...623..721P} Postman, M., Franx, M., Cross, N.~J.~G., et al.\ 2005, \apj, 623, 721
\bibitem[Postman et al.(2012)]{2012ApJ...756..159P} Postman, M., Lauer, T.~R., Donahue, M., et al.\ 2012, \apj, 756, 159 
\bibitem[Purcell et al.(2007)]{2007ApJ...666...20P} Purcell, C.~W., Bullock, J.~S., \& Zentner, A.~R.\ 2007, \apj, 666, 20

\bibitem[Rhode(2012)]{2012AJ....144..154R} Rhode, K.~L.\ 2012, \aj, 144, 154 
\bibitem[Rodriguez-Gomez et al.(2016)]{2016MNRAS.458.2371R} Rodriguez-Gomez, V., Pillepich, A., Sales, L.~V., et al.\ 2016, \mnras, 458, 2371 
\bibitem[Rudick et al.(2006)]{2006ApJ...648..936R} Rudick, C.~S., Mihos, J.~C., \& McBride, C.\ 2006, \apj, 648, 936 
\bibitem[Rudick et al.(2011)]{2011ApJ...732...48R} Rudick, C.~S., Mihos, J.~C., \& McBride, C.~K.\ 2011, \apj, 732, 48 
\bibitem[Rusli et al.(2013)]{2013AJ....146..160R} Rusli, S.~P., Erwin, P., Saglia, R.~P., et al.\ 2013, \aj, 146, 160

\bibitem[Schlafly \& Finkbeiner(2011)]{2011ApJ...737..103S} Schlafly, E.~F., \& Finkbeiner, D.~P.\ 2011, \apj, 737, 103
\bibitem[Schombert(1988)]{1988ApJ...328..475S} Schombert, J.~M.\ 1988, \apj, 328, 475
\bibitem[Seigar et al.(2007)]{2007MNRAS.378.1575S} Seigar, M.~S., Graham, A.~W., \& Jerjen, H.\ 2007, \mnras, 378, 1575

\bibitem[Sereno et al.(2013)]{2013MNRAS.428.2241S} Sereno, M., Ettori, S., Umetsu, K., \& Baldi, A.\ 2013, \mnras, 428, 2241

\bibitem[Sersic(1968)]{1968adga.book.....S} S\'ersic, J.~L.\ 1968, Atlas de Galaxias Australes (Cordoba: Observatorio Astronomico)
% Galaxy Zoo: the interplay of quenching mechanisms in the group environment
\bibitem[Smethurst et al.(2017)]{2017MNRAS.469.3670S} Smethurst, R.~J., Lintott, C.~J., Bamford, S.~P., et al.\ 2017, \mnras, 469, 3670  
\bibitem[Smith et al.(2013)]{2013MNRAS.429.1066S} Smith, R., S{\'a}nchez-Janssen, R., Fellhauer, M., et al.\ 2013, \mnras, 429, 1066 
\bibitem[Spitler \& Forbes(2009)]{2009MNRAS.392L...1S} Spitler, L.~R., \& Forbes, D.~A.\ 2009, \mnras, 392, L1
\bibitem[Spitzer \& Baade(1951)]{1951ApJ...113..413S} Spitzer, L., Jr., \& Baade, W.\ 1951, \apj, 113, 413
\bibitem[Stetson(1987)]{1987PASP...99..191S} Stetson, P.~B.\ 1987, \pasp, 99, 191 

% mass quenching for centrals; environmental quenching for satellites:
\bibitem[Tal et al.(2014)]{2014ApJ...789..164T} Tal, T., Dekel, A., Oesch, P., et al.\ 2014, \apj, 789, 164 
\bibitem[Tanaka et al.(2005)]{2005MNRAS.362..268T} Tanaka, M., Kodama, T., Arimoto, N., et al.\ 2005, \mnras, 362, 268
\bibitem[Thomas et al.(2016)]{2016Natur.532..340T} Thomas, J., Ma, C.-P., McConnell, N.~J., et al.\ 2016, \nat, 532, 340
\bibitem[Trujillo et al.(2004)]{2004AJ....127.1917T} Trujillo, I., Erwin, P., Asensio Ramos, A., \& Graham, A.~W.\ 2004, \aj, 127, 1917 

\bibitem[van den Bergh(1958)]{1958Obs....78...85V} van den Bergh, S.\ 1958, The Observatory, 78, 85
  
% The importance of satellite quenching for the build-up of the red sequence of present-day galaxies   
\bibitem[van den Bosch et al.(2008)]{2008MNRAS.387...79V} van den Bosch, F.~C., Aquino, D., Yang, X., et al.\ 2008, \mnras, 387, 79  
  
% The Physical Origins of the Morphology-Density Relation: Evidence for Gas Stripping from the Sloan Digital Sky Survey   
\bibitem[van der Wel et al.(2010)]{2010ApJ...714.1779V} van der Wel, A., Bell, E.~F.,
  Holden, B.~P., Skibba, R.~A., \& Rix, H.-W.\ 2010, \apj, 714, 1779   
\bibitem[Veale et al.(2017a)]{2017MNRAS.471.1428V} Veale, M., Ma, C.-P., Greene, J.~E., et al.\ 2017a, \mnras, 471, 1428 
\bibitem[Veale et al.(2017b)]{2017arXiv170800870V} Veale, M., Ma, C.-P., Greene, J.~E., et al.\ 2017b, arXiv:1708.00870 
%
\bibitem[V{\'{\i}}lchez-G{\'o}mez(1999)]{1999ASPC..170..349V} V{\'{\i}}lchez-G{\'o}mez, R.\ 1999, The Low Surface Brightness Universe, 170, 349 
\bibitem[Vulcani et al.(2012)]{2012MNRAS.420.1481V} Vulcani, B., Poggianti, B.~M., Fasano, G., et al.\ 2012, \mnras, 420, 1481
\bibitem[West et al.(1995)]{1995ApJ...453L..77W} West, M.~J., Cote, P., Jones, C., Forman, W., \& Marzke, R.~O.\ 1995, \apjl, 453, L77
\bibitem[West et al.(2004)]{2004Natur.427...31W} West, M.~J., C{\^o}t{\'e}, P., Marzke, R.~O., \& Jord{\'a}n, A.\ 2004, \nat, 427, 31
\bibitem[West et al.(2011)]{2011A&A...528A.115W} West, M.~J., Jord{\'a}n, A., Blakeslee, J.~P., et al.\ 2011, \aap, 528, A115

% (group preprocessing)
% Galaxy evolution in groups and clusters: satellite star formation histories and quenching time-scales in a hierarchical Universe 
\bibitem[Wetzel et al.(2013)]{2013MNRAS.432..336W} Wetzel, A.~R., Tinker, J.~L., Conroy, C., \& van den Bosch, F.~C.\ 2013, \mnras, 432, 336  

\bibitem[White(1987)]{1987MNRAS.227..185W} White, R.~E., III 1987, \mnras, 227, 185 
  
\bibitem[Yahagi \& Bekki(2005)]{2005MNRAS.364L..86Y} Yahagi, H., \& Bekki, K.\ 2005, \mnras, 364, L86
\bibitem[Zwicky \& Humason(1964)]{1964ApJ...139..269Z} Zwicky, F., \& Humason, M.~L.\ 1964, \apj, 139, 269 

\end{thebibliography}
\end{document}